\documentclass[aps,prx,twocolumn,showpacs,eqsecnum,superscriptaddress,floatfix,10pt]{revtex4-2}
\usepackage{amsmath,amssymb,amsfonts,stmaryrd,wasysym,graphicx,multirow,color,textcomp,subfigure,xcolor}
\usepackage{url,booktabs}
\usepackage[colorlinks=true,citecolor=blue,urlcolor=blue]{hyperref}
\usepackage{float}
\usepackage{color,soul} 
\usepackage{bm}
\usepackage[normalem]{ulem}

\def\be{\begin{equation}}
	\def\ee{\end{equation}}
\def\bea{\begin{eqnarray}}
	\def\eea{\end{eqnarray}}

\newcommand{\vv}[1]{{\boldsymbol #1}}

\begin{document}
	
	\title{Theory of Moir\'e Magnets and Topological Magnons \\: Applications to Twisted Bilayer CrI$_3$ }
	\author{Kyoung-Min Kim}
	\affiliation{Center for Theoretical Physics of Complex Systems, Institute for Basic Science (IBS) Daejeon 34126, Republic of Korea}
	\author{Do Hoon Kiem}
	\affiliation{Department of Physics, KAIST, Daejeon 34141, Republic of Korea}
	\author{Grigory Bednik}
	\affiliation{Center for Theoretical Physics of Complex Systems, Institute for Basic Science (IBS) Daejeon 34126, Republic of Korea}
	\author{Myung Joon Han}
	\email{mj.han@kaist.ac.kr}
	\affiliation{Department of Physics, KAIST, Daejeon 34141, Republic of Korea}
	\author{Moon Jip Park}
	\email{moonjippark@ibs.re.kr}
	\affiliation{Center for Theoretical Physics of Complex Systems, Institute for Basic Science (IBS) Daejeon 34126, Republic of Korea}
	\begin{abstract}
		We develop a comprehensive theory of twisted bilayer magnetism. Starting from the first-principles calculations of two-dimensional honeycomb magnet CrI$_3$, we construct the generic spin models that represent a broad class of twisted bilayer magnetic systems. Using  Monte-Carlo method, we discover a variety of non-collinear magnetic orders and topological magnons that have been overlooked in the previous theoretical and experimental studies. As a function of the twist angle, the collinear magnetic order undergoes the phase transitions to the non-collinear order and the magnetic domain phase. In the magnetic domain phase, we find that the spatially varying interlayer coupling produces the magnetic skyrmions even in the absence of the Dzyaloshinskii–Moriya interactions. In addition, we describe the critical phenomena of the magnetic phase transitions by constructing the field theoretical model of the moir\'e magnet. Our continuum model well-explains the nature of the phase transitions observed in the numerical simulations. Finally, we classify the topological properties of the magnon excitations. The magnons in each phases are characterized by the distinct mass gaps with different physical origins. In the collinear ferromagnetic order, the higher-order topological magnonic insulator phase occurs. It serves as a unique example of the higher-order topological phase in magnonic system, since it does not require non-collinear order or asymmetric form of the interactions. In the magnetic domain phases, the magnons are localized along the domain wall and form one-dimensional topological edge mode. As the closed domain walls deform to a open network, the confined edge mode extends to form a network model of the topological magnons. 
	\end{abstract}
	
	\maketitle
	
	\section{Introduction}
	
	After the discovery of superconductivity and correlated insulator phases in twisted bilayer graphene \cite{Cao2018,Cao20182}, engineering moir\'e superlattices now becomes a novel platform for designing exotic material properties and correlated phases of two-dimensional materials \cite{Andrei2020,Can2021,PhysRevLett.99.256802,PhysRevB.82.121407,Bistritzer12233,Wang2020,Fu2020,PhysRevB.102.180304,PhysRevLett.126.223601,PhysRevLett.126.136101,doi:10.1126/science.aav1910}. One of the promising routes to generalize the moir\'e superlattice engineering is to consider the twisted bilayer of two-dimensional magnets. Understanding the spin systems in the moir\'e superlattice is a challenging problem since the single unit cell accompanies a large number of the spin degrees of freedom. In this regard, there have been both experimental and theoretical efforts in pursuit of understanding moir\'e magnets. The emergence of the non-collinear order and magnetic domain wall has been experimentally observed in twisted bilayer of chromium triiodide (CrI$_3$) using nitrogen-vacancy centers magnetometer \cite{doi:10.1126/science.abj7478} and magnetic circular dichroism \cite{Xu2022}. Theoretical works have been also proposed using the effective continuum model description \cite{Hejazi10721,Akram2021}, monolayer-substrate model, which assumes the fixed spins on the substrate layer \cite{PhysRevResearch.3.013027,PhysRevB.103.L140406,doi:10.1021/acs.nanolett.8b03315}, and the network model of untwisted domains \cite{PhysRevLett.125.247201}.
	
	Although there have been several studies in the theory of moir\'e magnetism, most of the previous works have relied on either continuum approximation or the effective model, which ignores the mutual interactions between the layers. Here, we provide comprehensive theory of the moir\'e magnetism by fully taking account of the lattice structure and mutual spin interactions between the two layers. Our study offers important guidance in future experiments of moir\'e magnetism by providing the following theoretical breakthroughs: \textbf{(i)} In our model, the mutual spin interactions between the layers are fully taken into account in the Monte-Carlo simulations. The resulting magnetic ground states are different from the previous predictions in substrate models. \textbf{(ii)} The implementation of the full lattice model explains the magnetic structures beyond the small twist angle regime, where the continuum models have focused on. \textbf{(iii)} We provide the topological band theory of the magnonic excitations, based on the full lattice model and magnetic symmetry groups.
	
	The main goal of this work is to unveil the non-collinear magnetic orders of twisted bilayer magnet CrI$_3$ and provide a general theoretical framework that can be extended to various two-dimensional magnetic materials with different magnetic interactions and symmetries. By controlling the twist angle, we show that the collinear magnetic order in the aligned bilayer undergoes the phase transitions to the non-collinear domain, magnetic domain, and skyrmion domain phases. The developement of the non-collinear magnetic order is based on the geometric property of a moir\'e superlattice that different local stacking regions coexist within a single moir\'e unit cell. In particular, our theory explains the recent experimental observation of the magnetic domains of CrI$_3$ \cite{doi:10.1126/science.abj7478}, and therefore our comprehensive analysis will provide useful guidance for future experiments. Furthermore, by considering the general spin model with different sets of the spin interactions and anisotropies, we explicitly show that our theory of non-collinear phase transition is applicable to other magnetic materials such as XXZ-type and Ising type magnets.
	
\begin{figure*}[t]
		\includegraphics[width=1\textwidth]{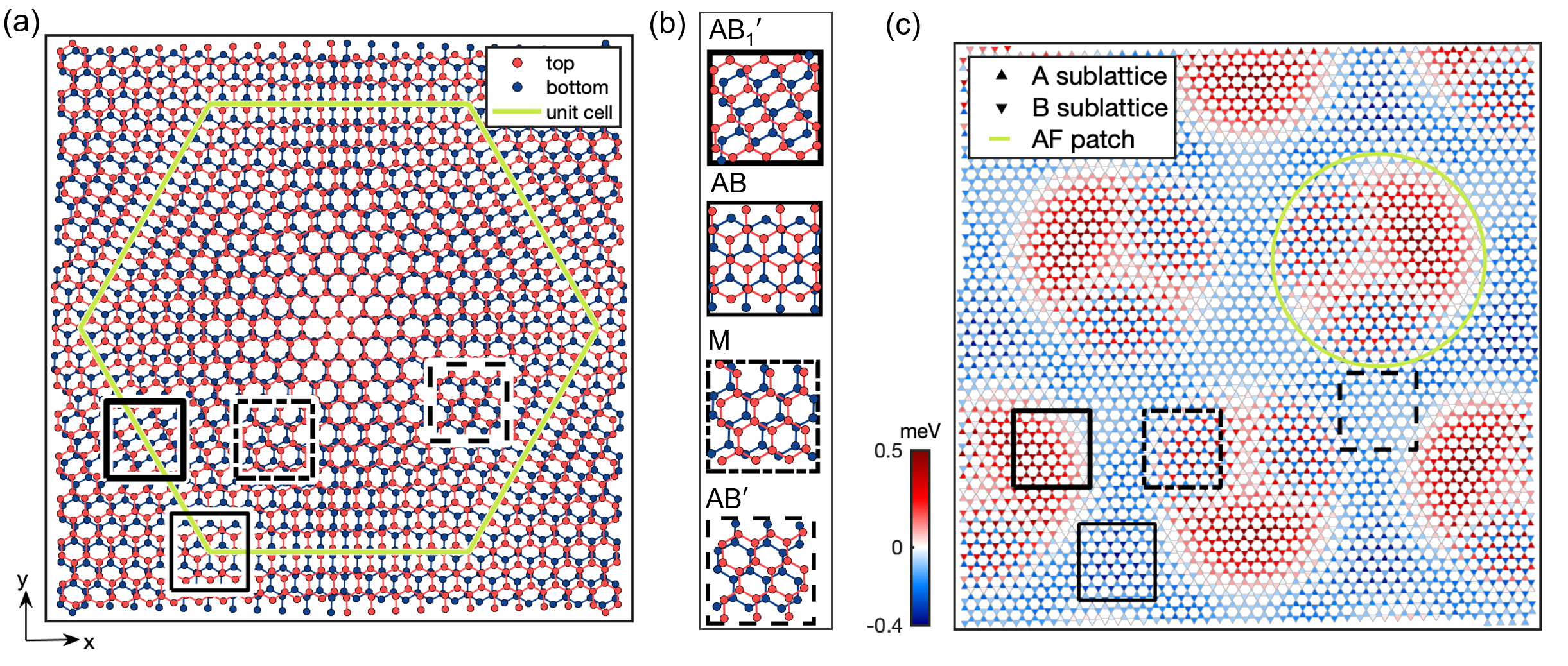}
		\caption{ (a) An example of moir\'e superlattice unit cell (green boundary) of CrI\textsubscript 3 with a twist angle 3.15\textdegree [$(m,n)=(10,11)$]. Each site represents a chromium atom, which carries the local magnetic moment $S=3/2$. Different local stacking structures coexist inside a single unit cell. (b) Magnified view of the local stacking structures. In the aligned bilayer CrI\textsubscript 3, AB' and metastable AB phases are stabilized and exhibits the interlayer antiferromagnetism and the interlayer ferromagnetism respectively. $\textrm{AB}'_1$ and M phase represent the unstable local stacking structure in the bilayer CrI\textsubscript 3, where the antiferromagnetic and the ferromagnetic couplings become maximal respectively. (c) Interlayer interactions in twisted bilayer CrI\textsubscript 3. The effective exchange field ($J_\perp$) acting on the spins on the top layer at the twist angle  $\theta=1.61^\circ[(m,n)=(20,21)]$. The triangles ($\triangle$) and the inverted triangles ($\triangledown$) represent the spins on A and B sublattices of the top layer respectively. The red (blue) color indicates the interlayer antiferromagnetic (ferromagnetic) couplings respectively.   }
		\label{fig:lattice}
	\end{figure*}

	Chromium trihalide family represented by CrI$_3$ has gathered great interests for the controllable two-dimensional magnetism \cite{Huang2017,doi:10.1021/acs.nanolett.8b03321,Jiang2018,Niu2020,Li2019,Wang_2011,SORIANO2019113662,Sun2019,PhysRevMaterials.3.031001,doi:10.1126/sciadv.1603113,Seyler2018,Jiang20182,Huang2018,doi:10.1126/science.aar3617,Wang_2011,PhysRevB.91.235425,doi:10.1021/acs.nanolett.8b01125,doi:10.1021/acs.nanolett.8b01125,C8NR03230K,PhysRevB.99.144401,SORIANO2019113662,Seyler2018,Jiang20182,Huang2018,doi:10.1126/science.aar4851,Wang2018,Kim2018,Song2019,PhysRevLett.121.067701}. The bilayer CrI$_3$ exhibits the magnetic phase transitions from the interlayer ferromagnetism (FM) to the interlayer antiferromagnet (AFM) accompanied by the structural phase transitions of stacking patterns \cite{Huang2017,doi:10.1021/acs.nanolett.8b03321,PhysRevMaterials.3.031001}. Such stacking dependent interlayer coupling poses an interesting physical scenario when the twisted bilayer is considered. In turn, twisted bilayer of the honeycomb magnet consists of local stacking patches, each of which interacts ferromagnetically (AB stacking) and antiferromagnetically (AB' stacking) simulatneously [See Fig. \ref{fig:lattice}]. The coexistence of the ferromagnetic and antiferromagnetic interlayer coupling generates the energetic competition between the collinear and the non-collinear magnetic ground states. In general, van der Waals magnets have weak interlayer couplings compared to the intralayer couplings. Nevertheless, we show that the sufficiently small twist angle stabilizes the non-collinear magnetic ground states, even when the strength of the interlayer coupling is arbitrarily small.
	
	Another important hallmark of honeycomb magnet is the topological magnons associated with Dirac magnons \cite{PhysRevLett.104.066403,PhysRevX.8.041028,Owerre_2016,doi:10.1063/1.4959815,PhysRevB.100.144401,mook2020interactionstabilized,PhysRevB.87.174427,PhysRevB.3.157,PhysRevLett.119.247202,PhysRevLett.123.227202}. Recently inelastic neutron scattering experiment explicitly observed the topological mass gap of the Dirac magnons in the bulk sample of CrI$_3$ \cite{PhysRevX.8.041028}. In general, the magnetic ground state and the associated band topology of the magnons are highly dependent on the form of the interlayer couplings \cite{PhysRevB.104.L060401}. Here, we show that each of the magnetic phases we uncovered is characterized by massive Dirac magnon excitations with different physical origins. We find various topological magnonic phases depending on the underlying symmetries and ground states. In the collinear magnetic order, we find the higher-order topological magnonic insulator (HOTMI) phase, in which the two-dimensional bulk topology characterizes the zero-dimensional corner magnon excitations \cite{Sil_2020,PhysRevLett.125.207204,PhysRevB.104.024406}. In contrast to the previous proposals, we show that this HOTMI can also be well-stabilized without the aid of non-collinear orders or asymmetric interaction. In the magnetic domain phases, the interior and the exterior of the domain have different Chern numbers, where the domain wall harbors a confined one-dimensional topological edge mode excitations. Due to the confinement of the domain, the edge mode manifests as the flat band with the macroscopic degeneracy. As the geometry of the magnetic domains varies and the domain walls are connected each other, the confined edge mode extends to form a network model of the topological magnons.

	This paper is organized as follows. In Sec. \ref{sec:model}, we consider the generic spin model of the twisted bilayer magnets. Based on the first-principles calculations, the effective spin model of twisted bilayer CrI$_3$ is constructed. In Sec. \ref{sec:MonteCarlo}, using Monte Carlo simulation, we clarify the magnetic phase transitions to the non-collinear orders as a function of the twist angle. In the small twist angles, we find the emergence of the non-collinear domain, magnetic domain, and skyrmion phases respectively. In particular, the skyrmion phase is stabilized in the absence of the Dzyaloshinskii–Moriya interactions (DMI) due to the peculiar spatial distributions of the interlayer couplings. We compare our results with previous experimental results. In Sec. \ref{sec:ContinuumModel}, we construct the generic continuum field theoretical model and compare it with the results obtained from the numerical Monte Carlo simulation. In Sec. \ref{sec:magnon}, we use the linear spin-wave theory to describe the magnon excitations of each phase. We show that the distinct topological invariants characterize both collinear and non-collinear order found in Sec \ref{sec:MonteCarlo}. In the collinear phase, the higher-order topological magnonic insulator phase is found. In the non-collinear phase, the first-order topology manifests as the edge modes localized on the domain walls. In Sec. \ref{sec:Conclusion}, we conclude our study by providing future research directions with the discussion of various van der Waals magnetic materials.
	
	\subsection{Lattice structures and symmetries of moir\'e superlattice} \label{sec:lattice_structure}

	Starting from AA-stacked honeycomb bilayer of CrI\textsubscript{3} (point group D\textsubscript{3d}), the rotation along one of the hexagonal center [see Fig.~\ref{fig:lattice} (a)] and the subsequent lateral shift of one layer with respect to the other generate all possible commensurate moir\'e superlattice structures. The rotation is characterized by the commensurate twist angles, which is given as $\theta_{m,n} =\arccos\big[\frac{1}{2}\frac{m^2+n^2+4mn}{m^2+n^2+mn}\big]$, where $m$ and $n$ are coprime integers \cite{PhysRevB.99.195455, PhysRevB.87.205404}. The twist results in an enlarged moir\'e unit cell with the lattice length $L=a_0\sqrt{m^2+n^2+mn}=\frac{|m-n|a_0}{\sin \theta_{m,n}/2}$ where $a_0$ is the monolayer honeycomb lattice length. The area of the moir\'e unit cell has general tendency to diverge $A\sim 1/\theta^2$ ($L\gg a_0$) in the small angle limit of $\theta\rightarrow0$.
	
	The subsequent lateral shift by a vector, $\mathbf{d}_{\textrm{shift}}$, does not change the size of the moir\'e unit cell, but it further lowers the global point group symmetry. We classify the lateraly shifted moir\'e superlattices of CrI\textsubscript{3} based on their point group symmetry [See Table \ref{tab:point_group_crystal} in Appendix \ref{Appsec:lattice}]. For example, when  $\mathbf{d}_{\textrm{shift}}=0$, the superlattice belongs to the point group D\textsubscript{3} by retaining the two rotational symmetry C\textsubscript{3z} and C\textsubscript{2y} along $z$ and $y$ axis, but the twist breaks S\textsubscript{6z} about the $z$ axis. This is different from the previously well-studied cases of the AA stacked twisted bilayer graphene (point group D\textsubscript{6}) \cite{PhysRevB.98.085435}, since the nonmagnetic iodine atoms break the two-dimensional rotation  C\textsubscript{2z}. An arbitrary lateral shift further breaks both C\textsubscript{3z} and C\textsubscript{2y}. However, there exist special lateral shifts that restore both symmetries by moving the rotational axis and centers to the other sites within a single moir\'e unit cell. The number of such symmetry restoring points increases as $\sim L^2/a_0^2$. Correspondingly, in the small twist angle limit, C\textsubscript{3z} and C\textsubscript{2y} are approximately preserved regardless of the lateral shift.

	\section{Spin Model in Moir\'e Superlattice}\label{sec:model}

	\subsection{First-principles calculations and effective Heisenberg spin model} \label{sec:spin_model}
	
	\begin{table}[b!]
		\begin{tabular}{cccccccc}
			\toprule
			$\textrm{J}_0 (\textrm{meV})$ &$\textrm{J}^\textrm{s}_1 ('') $ & $\textrm{J}^\textrm{c}_1 ('') $ & $l_0 (\textrm{\AA})$ & $l^\textrm{s}_1 ('')$ & $l^\textrm{c}_1  ('')$ & $\mathbf q^\textrm{s}_{1} (\textrm{\AA}^{-1})$ & $\mathbf q^\textrm{c}_{1} ('')$ \\
			\midrule
			$-0.1$ & $-0.5$ & $0.1$ & $0.1$ & $0.3$ & $0.6$ &   $(0.7,0)$ & $(1.73,1)$\\
			\bottomrule
		\end{tabular}
		\caption{The parameters of the interlayer Heisenberg interaction in Eq.~(\ref{eq:J_inter}) obtained from the first-principles calculation of CrI\textsubscript{3}.}
		\label{tab:J_inter_params1}
	\end{table}

	Spins at each site of the twisted bilayer [Fig.~\ref{fig:lattice} (a)] interact with each other through the intralayer and interlayer Heisenberg exchange couplings. The spin model can be generally represented as,
	\begin{equation} \label{eq:spinH}
		H = -\sum_{\langle i,j \rangle} J \mathbf S_i \cdot \mathbf S_j + \sum_{z_j=z_i+d}J^{\perp}_{ij} \mathbf S_i \cdot \mathbf S_{j} - D_z\sum_i (S_i^z)^2,
	\end{equation}
	where $J>0$ represents the constant ferromagnetic intralayer Heisenberg interactions for nearest-neighbor spins. $D_z$ is the single-ion anisotropy \cite{Lado_2017}. $J^{\perp}_{ij}$ represents interlayer Heisenberg interactions, which depends on the relative coordinate displacement between the spins. We derive the interlayer coupling $J^{\perp}_{ij}$ for CrI\textsubscript{3} from the first-principles calculation by considering the fractional lateral shifts from the AA-stacked structure. For each stacked structure, we obtain the interlayer coupling $J_{ij}^\perp$ from the magnetic force theorem (MFT) \cite{PhysRevMaterials.3.031001,D1NR02480A}. (See Appendix \ref{Appsec:model} for the detailed methods).

	\begin{figure*}
		\includegraphics[width=\textwidth]{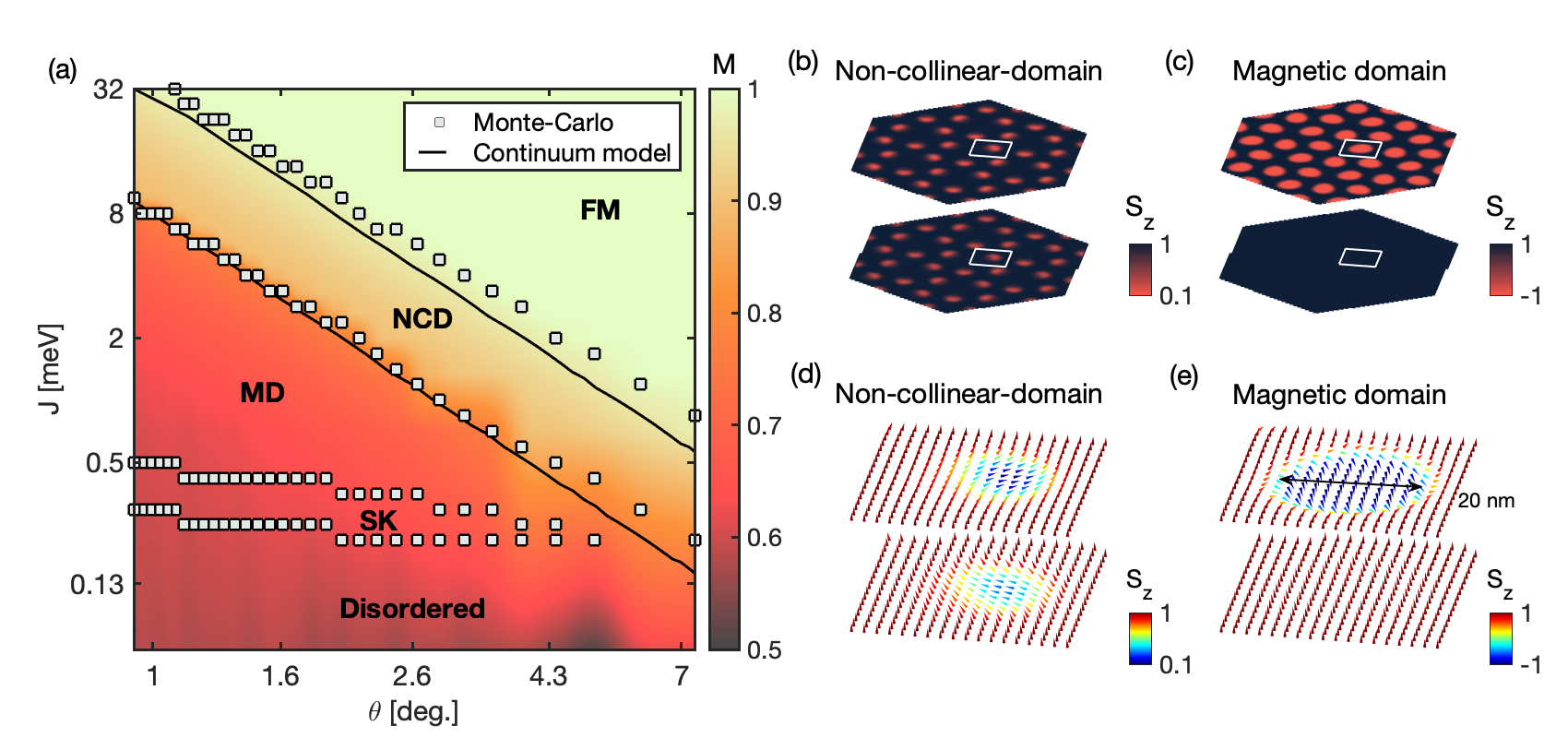}
		\caption{(a) The magnetic phase diagram showing the ferromagnetic order-parameter ($M$) as a function of the twist angle and the intralayer exchange coupling. Here, FM, NCD, MD, and SK denote the collinear ferromagnetic phase, the non-collinear domain phase, the magnetic-domain phase, and the skyrmion domain phase, respectively. (b)-(c) The examples of the magnetic structures for (b) NCD phase and (c) MD phase respectively. In the NCD phase, both layers exhibit the non-collinear spin tilting to the horizontal direction in the interior of the domain. In the MD phase, one layer exhibits the magnetic domain with $\pi$-rotation of the spins, while the other layer exhibits the nearly ferromagnetic ground state. (d)-(e) Magnified view of the spin texture of a single domain for (d) NCD phase and (e) MD phase respectively. In (b)-(e), the twist angle $\theta=0.93^\circ$ is used. In (b) and (d) ((c) and (e)), $J=11.3$ ($J=4$) meV is used. In all plots, $D_z=0.2$ meV is used. }
		\label{fig:MC}
	\end{figure*} 
	
	Based on the first-principles calculation, we generally write down the continuum form of the C$_{3z}$-rotational symmetric interlayer couplings, which is given as,
	
	\begin{align} \label{eq:J_inter}
		J^{\perp}_{ij} & = \textrm{J}_0 \exp\big(-|r_{ij}-d_0|/l_0\big)
		\nonumber \\
		&+\textrm{J}^\textrm{s}_1 \exp\big(-|r_{ij}-r_*|/l^\textrm{s}_1\big) \sum_{a=1}^{3}\sin(\mathbf q^\textrm{s}_{a} \cdot \mathbf r_\parallel) 
		\nonumber \\
		&+ \textrm{J}^\textrm{c}_1 \exp\big(-|r_{ij}-r_*|/l^\textrm{c}_1\big) \sum_{a=1}^{3}\cos(\mathbf q^\textrm{c}_{a} \cdot \mathbf r_\parallel),
	\end{align}
	where $\mathbf r_{\parallel}=(x_i-x_j,y_i-y_j)$ is the parallel displacement vector between the two spins at site $i$ and $j$. $d_0=6.7 \AA$ is the interlayer distance.  $r_{ij}=\sqrt{|\mathbf r_{\parallel}|^2+d_0^2}$. $r_* =7.3 \AA$ is the characteristic length scale. Due to the C\textsubscript{3z} symmetry, the wave vectors, $\mathbf q^\textrm{s}_{a}$ and $\mathbf q^\textrm{c}_{a}$, are related by 120$^\circ$ rotation as $\mathbf q^{\textrm{s,c}}_{1}=\textrm{R}(-2\pi/3) \mathbf q^{\textrm{s,c}}_{2}=\textrm{R}(-4\pi/3) \mathbf q^{\textrm{s,c}}_{3}$ where $\textrm{R}(\theta)$ is the  rotation matrix. The detailed values of the parameters are summarized in Table \ref{tab:J_inter_params1}. The first term in Eq. \eqref{eq:J_inter} is the simple ferromagnetic interactions, and the second and the third term characterize the oscillatory behaviors, which are responsible for the stacking dependent magnetism. In principle, the higher-order oscillatory terms can be included, but we find that the first-order term is enough to describe the case of CrI\textsubscript{3}. (See Appendix \ref{Appsec:model} for the comparison of the DFT result and the effective model.)
	
	Within a single moir\'e unit cell, the spatially varying local stacking pattern gives rise to the competing interlayer interactions within a unit cell. For instance, Fig. ~\ref{fig:lattice} (b) presents the local stacking structures. We find that AB, AB', AB$_1$', and M stacking structures simultaneously appear in which each local structure possesses either ferromagnetic or antiferromagnetic interlayer couplings. Correspondingly, the interlayer spin interaction shows the spatial variation. Fig.~\ref{fig:lattice} (c) shows the effective exchange field of the ferromagnetic bottom layer acting on the top layer through the interlayer coupling, which is defined as,
	\begin{equation} 
		\label{eq:J_inter_eff}
		J_\perp (\mathbf x_i) = \sum_{j\in \textrm{bottom layer}} J^\perp_{ij},
	\end{equation}
	where $\mathbf{x}_i$ represents the coordinate of the spins on the top layer. While the average interlayer coupling ($\sum_{i}J_\perp (\mathbf x_i)$) is found to be ferromagnetic, we find the well-defined local antiferromagnetic patches (red triangles inside green circles) on the ferromagnetic background regions (blue triangles). Furthermore, the antiferromagnetic patches can be divided into the three distinct regions depending on how the local field acts on different sublattices: (i) both A and B sublattices are antiferromagnetic, (ii) only A sublattice are antiferromagnetic, while B sublattice is ferromagnetic, and (iii) the opposite of the region (ii). This sublattice selective local field has not been found in previous studies. In the microscopic model, the origin of the sublattice selection is attributed to the iodine atoms that break the C$_{2z}$ symmetry, which exchanges A and B sublattice degree of the freedom. In the next sections, we show that the sublattice selective interaction plays a crucial role in determining the magnetic ground state.

	\begin{figure}
		\includegraphics[width=.5\textwidth]{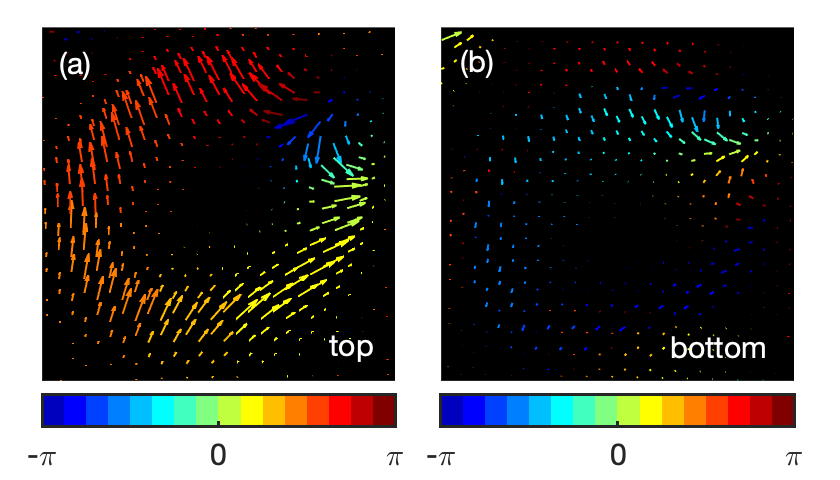}
		\caption{The horizontal spin texture of the skyrmion domain phase for (a) the top and (b) the bottom layer. The interior of the domain has $z$-directional interlayer antiferromagnetic order while the exterior of the domain has $z$-directional interlayer ferromagnetic order. }
		\label{fig:skyrmion}
	\end{figure} 
	
	\section{Monte Carlo Simulations : Emergence of Non-collinear Order} \label{Sec:montecarlo} \label{sec:MonteCarlo}

	\subsection{Heisenberg spin Model}
	
\begin{figure}
		\includegraphics[width=.5\textwidth]{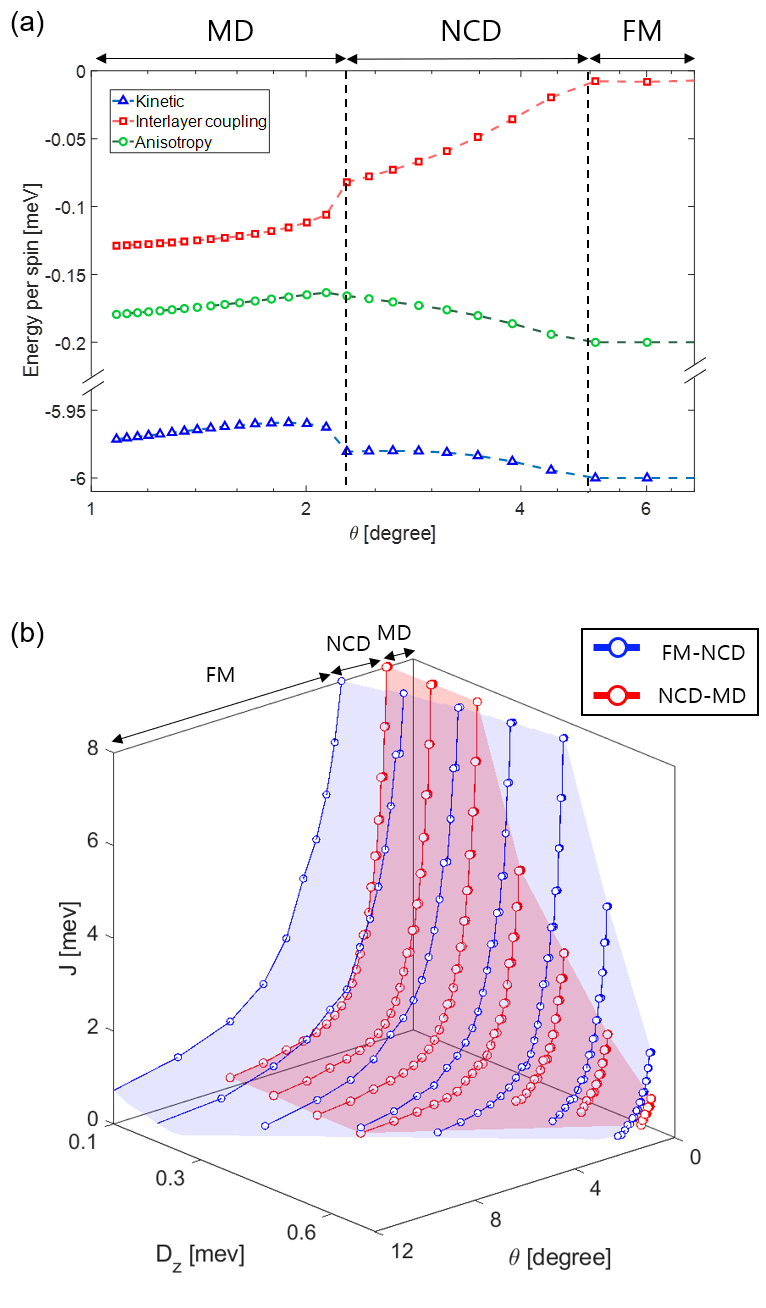}
		\caption{(a) The comparison of the kinetic energy ($-J \mathbf S_i \cdot \mathbf S_j $), the interlayer coupling energy ($J^{\perp}_{ij} \mathbf S_i \cdot \mathbf S_{j} $), and the single-ion anisotropy energy ($-D_z\sum_i (S_i^z)^2$) per spin as a function of the twist angle. (b) Three-dimensional magnetic phase diagram as a function of the intralayer coupling, twist angle, and single-ion anisotropy. As the single-ion anisotropy increases (approaching to the Ising spin limit), the non-collinear magnetic orders are suppressed.}
		\label{fig:global_phase_diagram}
	\end{figure}

	After constructing the microscopic spin model, we perform the classical Monte-Carlo simulations (See Appendix \ref{Appsec:MC} for the detailed methods). Figure ~\ref{fig:MC} (a) plots the ferromagnetic order parameter as a function of the nearest neighbor intralayer couplings and the twist angle, which is given as,
	\begin{equation} \label{eq:order_parameter}
		M = \frac{1}{N}  \sum_{i=1}^{N} \mathbf{S}_i   ,
	\end{equation}
	where $\mathbf{S}_i $ represents a spin vector at each site $i$, and $N$ is the number of spins within a single moir\'e unit cell. We first focus on the case of CrI$_3$, which corresponds to the intralayer coupling $J\approx 2.2 - 2.7 $meV \cite{Lado_2017,C5TC02840J}. In the large twist angle limit ($\theta \gtrsim 5^\circ$), the non-collinear orders cost a large gradient energy proportional to the intralayer coupling. [See Fig. \ref{fig:global_phase_diagram} (a) for the energy comparison of each phase.] We find that the interlayer ferromagnetic phase (FM), indicated by the saturation of the ferromagnetic order parameter, $M\approx 1$, is stabilized.
	
	Away from the FM phase, as the twist angle decreases ($2^\circ \lesssim\theta \lesssim 5^\circ$), the continuous decrease of the ferromagnetic order parameter $M$ is observed, which signals the magnetic phase transition from the ferromagnetic phase to non-collinear domain (NCD) phase. In the NCD phase, both layers show the spin tilting to the horizontal direction in the interior of the local antiferromagnetic patches (Fig.~\ref{fig:MC} (b)). Compared to the FM phase, the interlayer coupling energy density gain scales with the tilting angle ($\Phi_0$) while the intralayer gradient energy density cost rather scales with the rate of change of the tilting angle ($\Phi_0/R$). Therefore, in the small twist angle limit, the interlayer coupling energy dominates over the intralayer gradient energy, and as a result the NCD phase is favored over the FM phase. We emphasize that this non-collinear phase transition occurs in the sufficiently small twist angle even in the presence of the arbitrary weak strength of the interlayer couplings. [See the phase boundary of FM-NCD phases in Fig. \ref{fig:global_phase_diagram} (a)].

	As the twist angle is further lowered ($\theta \lesssim 2^\circ$), we find the additional non-collinear ordered phases distinguished from the NCD phase, which we refer to as magnetic domain (MD) phase. $\pi$-spin flip occurs in one of the layers across the domain wall [Fig.~\ref{fig:MC} (c)]. The interior of the AF local patches exhibits the interlayer antiferromagnetic domain, while the outside of the AF local patch shows the interlayer ferromagnetism. The MD phase occurs due to the presence of the small anisotropy, which energetically favors the polar-directional spin-flip ($\sim\pi$) in one of the layers rather than the two horizontal-directional spin tilting ($\lesssim\pi/2$) from both layers. In contrast, the large spin tilting of the MD phase costs more gradient energy than the case of the NCD phase. Thus, the MD phase is stabilized in the smaller angles than the NCD phase [See the phase boundary of MD-NCD phases in Fig. \ref{fig:global_phase_diagram} (a)].

	We identify two more magnetic ground states, as the intralayer coupling is decreased. In the small intralayer coupling limit ($J\rightarrow 0$), the magnetic ground state is disordered due to the loss of intralayer coherence. In the intermediate intralayer coupling strength between the MD phase and the disordered phase, the skyrmion (SK) magnetic domain phase is additionally distinguished from the MD phase. Along the one-dimensional domain wall, the orientation of the azimuthal spins winds through the equator in the Bloch sphere [See Fig. \ref{fig:skyrmion} (a)]. The spin texture ($\mathbf{S}= (\mathbf{S}_{\textrm{top}},\mathbf{S}_{\textrm{bot}})\in \textrm{S}^2\times \textrm{S}^2$) can be topologically classified by the two independent skyrmion numbers defined in each layer ($\pi_2(\textrm{S}^2\times \textrm{S}^2)=\mathbb{Z}\times \mathbb{Z}$), which is given as,
	\bea
	N_{\textrm{SK}}=\frac{1}{4\pi} \int_\mathcal{A} d^2 x \mathbf{S}_{\textrm{top(bot)}}\cdot (\partial_x \mathbf{S}_{\textrm{top(bot)}} \times \partial_y\mathbf{S}_{\textrm{top(bot)}}).
	\nonumber
	\\
	\eea
	where $\mathcal{A}$ represents the two dimensional area including the interior and the exterior of the domain. In the SK phase, one of the layer carries the non-trivial skyrmion number, while the spins of the other layer form nearly ferromagnet (trivial skyrmion number) [See Fig. \ref{fig:skyrmion} (b)].
	We note that the skyrmion phase occurs even in the absence of asymmetric interactions such as DMI, as we focus on the simple Heisenberg model. The absence of the inversion symmetry and the sublattice selective interlayer interaction discussed in Sec. \ref{sec:spin_model} are essential for the stabilization of the skyrmion phase since they effectively behave as the inversion breaking exchange field. The addition of the small intrinsic DMI along $z$ direction further enhances the stability of the SK phase.
	
	Finally, we classify the magnetic symmetry of each phase. The group of the symmetry operation that leaves the Heisenberg model in Eq. \eqref{eq:spinH} invariant is described by the spin-space group \cite{doi:10.1098/rspa.1966.0211,doi:10.1063/1.1708514,PhysRevB.105.064430}, where the symmetry operations of the spin space and the real space are effectively decoupled. The corresponding spin-space group, $G_s=G\times S$ is the direct product of the space group, $G$, which includes $C_{2y}$ and $C_{3z}$ with the lattice translations, and the spin group, $S=\langle\{U(1), \mathbb{Z}_2\}\rangle$, where $U(1)$ is the spin-rotation along $z$-axis and $\mathbb{Z}_2$ maps $S_z\rightarrow -S_z$. We classify the symmetry of the magnetic phases in Table \ref{tab:magnegtic_structure}. While the point group symmetries and spin rotations are broken, their product can be intact. For example, in the FM phase, $\mathbb{Z}_2 \times C_{2y}$ is preserved, while each of which is broken. Each magnetic phase is guaranteed to remain well-stabilized even in the presence of a small perturbation that preserves the underlying symmetries, summarized in Table \ref{tab:magnegtic_structure}. 
	
	\subsection{Ising and XY spin models}
	
	After we have established the magnetic phase diagram of the nearly Heisenberg spin models, we describes the scenario in the generic spin models.
	Fig. \ref{fig:global_phase_diagram} (b) presents the three-dimensional phase diagram as an additional function of the single-ion anisotropy. While we find the overall similar phase structure, the non-collinear order is strongly suppressed in the positive anisotropy region due to the additional energy cost. As a result, in the limit of the Ising spin ($D_z  \gg 1 $), all non-collinear orders are destabilized, and the FM phase survives regardless of the twist angles. On the other hand, in the opposite limit ($D_z \ll -1$), the spins effectively behave as XY type. Since NCD and MD phases are co-planar orders, the stabilities of these two non-collinear orders are unaffected compared to the case of the ideal Heisenberg model ($D_z = 0$). On the other hand, we find that the skyrmion phase, which is non-co-planar order, is quickly destabilized in the XY model limit. We remark that the qualitative features of our analysis are generally applicable in the generic spin systems in the presence of the stacking-dependent magnetism.

	\begin{table}[b!]
		\begin{tabular}{c|c|c|c|c|c|c|c}
			\toprule
			Phase & $U(1)$ & $\mathbb{Z}_{2}$ & $C_{2y}$ & $C_{3z}$ 
			& $\mathbb{Z}_2 \times C_{2y}$ & $\mathbb{Z}_{2y}\times C_{2y}$ & $\mathbb{Z}_{3z}\times C_{3z}$
			\\ 
			\midrule
			FM & $+$ & $-$ & $-$ & $+$ & $+$ & $-$ & $-$ \\
			NCD & $-$ & $-$ & $-$ & $-$ & $-$ & $+$ & $+$ \\
			MD & $-$ & $-$ & $-$ & $-$ & $-$ & $-$ & $+$  \\
			\bottomrule
		\end{tabular}
		\caption{Classification of magnetic structures in twisted bilayer honeycomb magnets based on the spin rotation and point group symmetry. $\mathbb{Z}_{2y}$ and $\mathbb{Z}_{3z}$ indicates the two-fold and the three-fold rotations of spins along $y$ and $z$ axis respectively. ``$+~(-)$" indicates that the corresponding symmetry is respected (broken).}
		\label{tab:magnegtic_structure}
	\end{table}
	
	\begin{figure*}[t]
		\includegraphics[width=\textwidth]{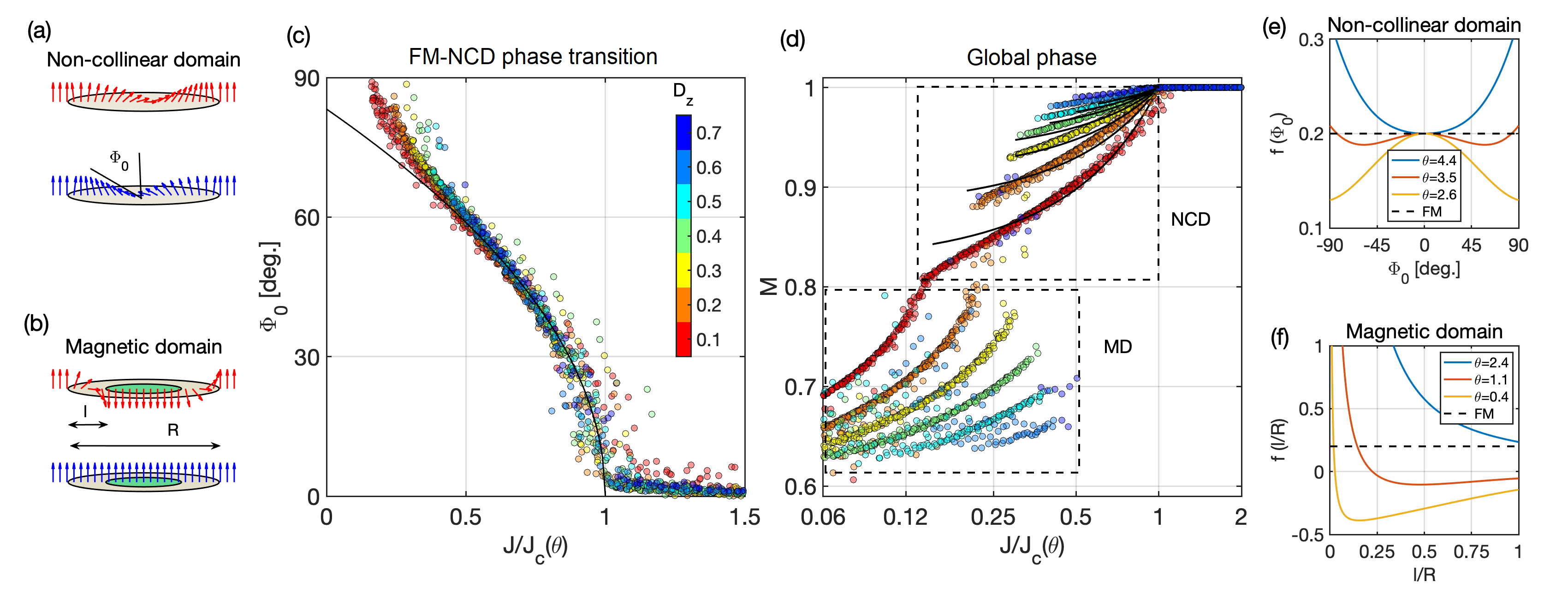}
		\caption{The result of continuum model in comparison to the Monte-Carlo simulation. Schematic illustration of the analytical solution for (a) the NCD and (b) the MD phases respectively. (c) The scaling analysis of the tilting angle, $\Phi_0$, near FM-NCD phase transition. The line and markers represents the results obtained from the continuum model and the Monte-Carlo simulations respectively. The FM-NCD phase transition is well-described by the Landau theory of the continuous phase transition. (d) The evolution of the ferromagnetic order parameter. The line and marker represent the results obtained from the continuum model and Monte-Carlo simulations. (e)-(f) The calculated free energy for different twist angles near (e) FM-NCD phase transition and (f) NCD-MD phase transition respectively. The free energy of the FM-NCD (NCD-MD) shows the second (first) order phase transitions.}
		\label{fig:ContinuumModel}
	\end{figure*} 
	
	\section{Continuum Model Description of Magnetic Phase Transitions}\label{sec:ContinuumModel}
	
	After establishing the magnetic phase diagram, we construct the low-energy effective field theory that describes the critical behaviors and the magnetic phase transitions. When the spatial variation of the local spin vector vary slow enough ($\nabla \mathbf{n}_{t,b}(\mathbf x)\ll 1$), the spin model in Eq. \eqref{eq:spinH} can be well-approximated by the following continuum free energy functional: 
	\bea 
	\label{eq:F}
	F[\mathbf{n}_{t},\mathbf{n}_{b}] & = \sum_{l=t,b} \int d^2\mathbf x \bigg\{ \frac{3a_0^2}{2} J [\nabla_\mathbf x \mathbf n_l(\mathbf x)]^2 - D_z [n_l^z (\mathbf x)]^2\bigg\} \nonumber \\
	& + \int d^2\mathbf x ~ \bar J_\perp ~ \mathbf n_t(\mathbf x) \cdot \mathbf n_b(\mathbf x),
	\eea
	where $\mathbf n_l(\mathbf x)$ is a normalized spin vector for $l$-th layer.
	$J$ and $D_z$ are the same magnetic interactions given in Eq.~(\ref{eq:spinH}), and $\bar J_\perp$ is the coarse grained value of interlayer coupling, $J^\perp_{ij}$. Using this free energy functional, we now describe the field theoretical model of the magnetic phase transitions of each magnetic phases.

	\begin{figure*}
		\includegraphics[width=\textwidth]{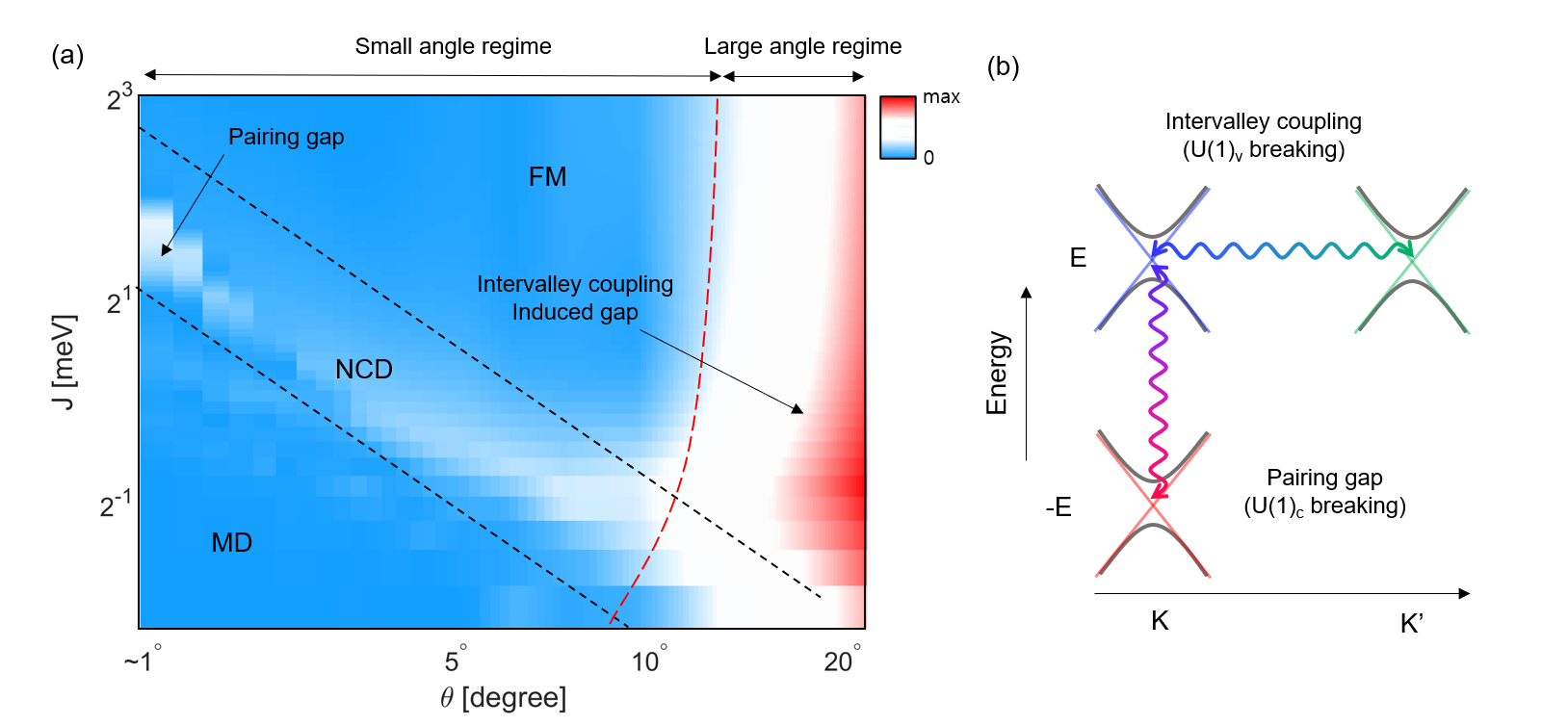}
		\caption{(a) Magnon gap as a function of intralayer coupling strength and the twist angle. The small angle regime of the FM phase is shown to be gapless, which is analogous to the electron spectrum of graphene. In the large angle regime of the FM phase, the short wave-length interlayer exchange interaction induces a finite coupling between the Dirac magnons in the different valleys, and it gives rise to a mass gap. In the NCD phase, the spin tilting breaks the $U(1)$ spin rotation symmetry. In the magnon spectrum, the lack of $U(1)$  manifests as the finite pairing gap (white region), distinct from the intervalley coupling induced gap. (b) Schematic illustration of the physical origin of the magnon gap. In the large angle FM phase, $U(1)_v$ symmetry breaking due to the intervalley coupling (Dirac cones at $\textrm{K}$ and $\textrm{K}'$) induces the magnon gap. In the NCD phase, the finite pairing breaks $U(1)_c$ symmetry and induces the magnon gap. }
		\label{fig:magnon}
	\end{figure*} 
	
	\subsection{Continuous Phase Transition : FM and NCD phases} \label{sec:F_NC}
	
	The NCD phase is characterized by the antiferromagnetic spin tilting to the azimuthal direction [Fig.~\ref{fig:ContinuumModel} (a)]. Without loss of generality, the spin tilting can be expressed as,
	\bea \label{eq:spin_conf_NC}
	\nonumber
	\mathbf n_t & = \big(\sin \Phi_{\textrm{t}} , 0, \cos \Phi_{\textrm{t}} \big), \\
	\mathbf n_b & = \big(-\sin \Phi_{\textrm{t}}, 0, \cos \Phi_{\textrm{t}} \big),
	\eea
	where the dependence of the tilting angle of the spin, $\Phi_{\textrm{t}}(\rho)=\Phi_0(1-\rho/R)$, is evaluated up to the linear-order as a function of the distance, $\rho$, from the center of the domain. $R$ is the size of the non-collinear magnetic domain. 
	Inserting the above ansatz of the NCD phase into Eq.~(\ref{eq:F}), we obtain the expression of the free energy functional up to the quartic order as a function of the tilting angle $\Phi_0$ as,

	\bea\label{eq:F_NCD}
	F[\Phi_0] &=& N_{\textrm{ncd}}(\theta) (\bar J_\perp - 2D_z) 
	\\
	\nonumber
	&+& \frac{a}{2}[J -J_c(\theta)]\Phi_0^2 + \frac{b}{4} J_c(\theta) \Phi_0^4+ \mathcal O (\Phi_0^6).
	\eea
	$N_{\textrm{ncd}}(\theta)$ is the number of tilted spins in the non-collinear magnetic domain, which scales as a function of the twist angle as, $N_{\textrm{ncd}}(\theta) \sim 1 /\theta^2$. $J_c(\theta)$ is the characteristic energy scale given as,
	$J_c(\theta) = c N_{\textrm{ncd}}(\theta) (\bar J_\perp - D_z),$
	and $a,b,c$ are $O(1)$ constant (See Appendix \ref{Appsec:F_NCD} for the details of the analytical calculation). We consider the Landau theory of the second-order phase transition as a function of $\Phi_0$. As the twist angle decreases from large value, the sign of the coefficient of the quadratic term changes from positive to negative, which develops the non-trivial free energy minimum ($\Phi_0\neq 0$) [See Fig.~\ref{fig:ContinuumModel} (e)]. As a result, the continuous phase transition occurs from FM to NCD phase. Near the phase transition boundary, the local minimum $\Phi_0$ follows the behavior of the order parameter in the standard second order phase transition as,
	\begin{equation} \label{eq:Phi_NC}
		\Phi_0 = \pm \sqrt{(a/b)\big[1-J/J_c(\theta)\big]}.
	\end{equation}
	Fig.~\ref{fig:ContinuumModel} (c) explicitly compares the tilting angle in the continuum theory model and the Monte-Carlo simulation. We clearly observes the behavior of the second order phase transition from the numerical result obtained from the Monte-Carlo simulation. We conclude that the FM-NCD phase transition falls into the category of standard second-order phase transition of the Landau theory.

	\subsection{First-order Phase Transition : NCD and MD phases} \label{sec:free_energy_DW}
	
	In contrast to the NCD phase, the MD phase has the spin flipping in one of the layer, and the other layer is nearly ferromagnetically ordered [see Fig.~\ref{fig:ContinuumModel} (b)]. The spin configuration of the MD phase for each layer is expressed as,
	\begin{align}
		\mathbf n_t & = \big(\sin \Phi_\textrm{t} , 0 ,\cos \Phi_\textrm{t}\big), \nonumber \\
		\mathbf n_b & = \big(0 , 0 , 1 \big),
	\end{align}
	where $\Phi_\textrm{t}(\rho) = \pi$ represents $\pi$-spin flip of the top layer in the interior of the domain for $0 \leq \rho \leq R-l$. $\Phi_\textrm{t}(\rho) = \pi (R-\rho)/l$ represent the continuous spin flip in the domain wall region for $R-l < \rho \leq R$, where $R$ and $l$ are the size of the magnetic domain and the width of the domain wall, respectively. Inserting this expression into Eq.~(\ref{eq:F}), we obtain the corresponding expression of the free energy for the MD phase as a function of the width of the domain wall $l$ as
	\begin{align} \label{eq:F_MD}
		F[l] & = \frac{aJ\pi^2}{4}\bigg(\frac{2R}{l}-1\bigg) -D_z N_{md}(\theta)\Bigg[\frac{3}{2} + \frac{1}{2}\bigg(1-\frac{l}{R}\bigg)^2\Bigg]
		\nonumber \\ 
		&
		-\bar J_\perp N_{md}\big(\theta \big) \Bigg[\bigg(1-\frac{l}{R} \bigg)^2 -\frac{4}{\pi^2}\bigg(\frac{l}{ R}\bigg)^2\Bigg],
	\end{align}
	where $N_{md}(\theta)$ is the number of spins in the magnetic domain, which scales as a function of the twist angle as $N_{md}(\theta) \approx \frac{1}{ 3\theta^2 }$. In the large twist angle, the free energy has only a trivial minimum with $l=R$, i.e. a magnetic state without magnetic domains. When $\theta$ becomes small, however, $F$ develops a nontrivial minimum for $l < R$ [See Fig.~\ref{fig:ContinuumModel} (f)], which corresponds to the stable MD phase. If this minimum has a lower energy than NCD state, the first-order phase transition from the NCD phase to the MD phase occurs [See Fig.~\ref{fig:ContinuumModel} (d)].  (See Appendix \ref{Appsec:F_MD} for the details of the analytical calculation and the case when spin tilting is introduced in the bottom layer).

	\section{Topological Magnons : Linear Spin Wave Theory}\label{sec:magnon}
	
	The non-collinearity of the magnetic ground state plays a crucial role in the description of the magnon excitations. Especially, the Dirac magnons that exists in the honeycomb magnet is the subject of great interest for the realization of the topological phases. In this section, we now turn our attention to the magnon excitations. We construct the linear spin wave theory of each magnetic phases we uncovered in Sec. \ref{sec:MonteCarlo}. (See Appendix \ref{Appsec:magnon1} for the detailed methods). 
	
	\begin{figure}
		\includegraphics[width=0.5\textwidth]{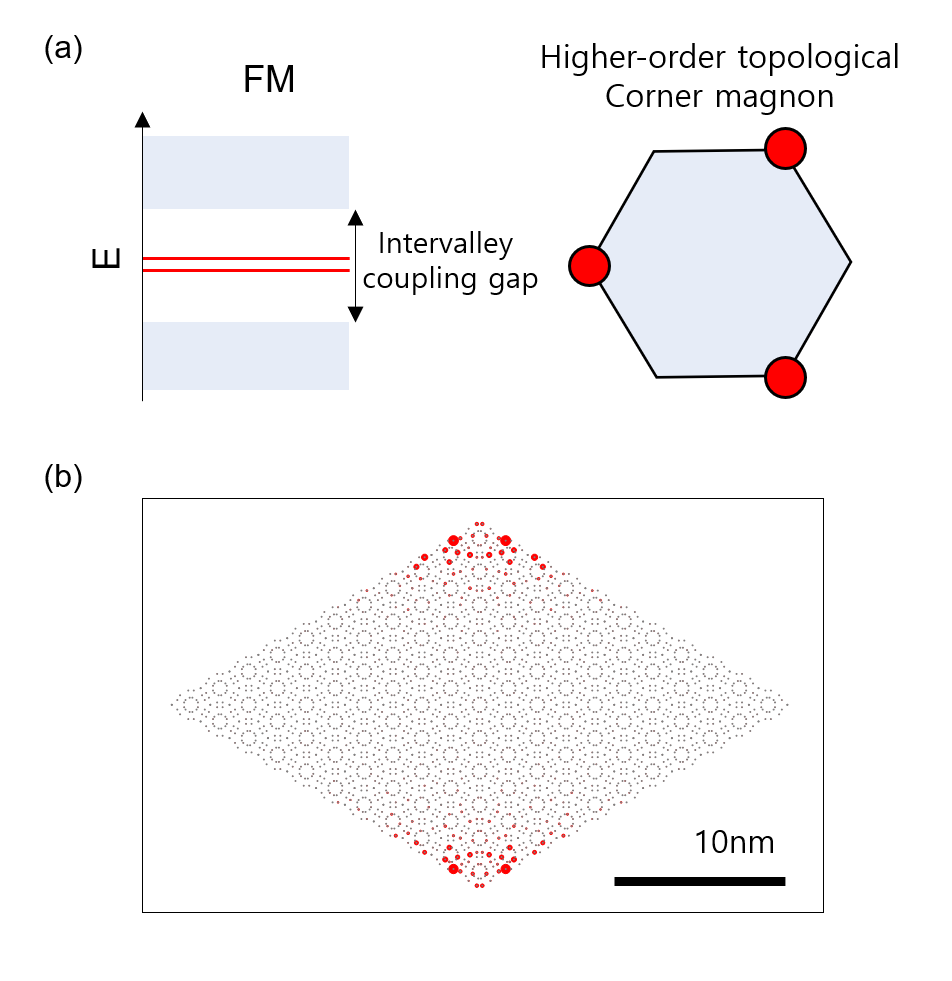}
		\caption{(a) Schematic figure of the topological corner modes. In the large angle FM phase, when the Dirac cone opens up the finite mass gap by the intervalley coupling, the HOTMI is realized with the topological corner mode. (b) Numerically calculated topological corner mode at the twist angle, $\theta=21.78^\circ$ [(m,n)=(1,2)]. Red dots indicate the amplitude of the topological corner wave functions. }
		\label{fig:magnoncorner}
	\end{figure} 
	
	\subsection{Case of collinear FM : HOTMI phase}
	
	In the collinear FM phase, the $U(1)$-spin rotation symmetry along the collinear axis exists ($U(1)_{c}$), and it allows to decouple the Hilbert space of the magnon Hamiltonian into the particle-hole sectors as, 
	\bea
	H_{\textrm{BdG}}(\mathbf{k})=H_{+}\oplus H_{-},
	\eea
	where each sector is composed of the single-particle magnon excitations carrying spin $\pm1$ respect to the collinear axis respectively. Each subsector can be diagonalized as $H_{\pm}=\pm U D U^\dagger |_{\pm}$. The corresponding eigenstates satisfy the unitary condition, $U U^\dagger|_{\pm}=U^\dagger  U|_{\pm}=\textrm{I}$, as fermionic eigenstates do. Thus, the system described by the equivalent tight-binding model and the topological classification is equivalent to the (spinless) electronic systems \cite{PhysRevB.104.L060401}. 
	
	In the absence of the interlayer coupling, each layer with the collinear FM phase exhibits the gapless Dirac magnons at $\textrm{K}$ and $\textrm{K}'$ points, which is analogous to electronic band structure of the monolayer graphene. Accordingly, the magnon spectrum in each valley possesses the linearized Dirac Hamiltonian with the four fold degeneracy (the total four Dirac cones from top and bottom layers). Under a proper rotation of the basis, the corresponding magnon Hamiltonian can be expressed as \cite{Bistritzer12233,PhysRevB.99.195455,PhysRevB.98.085435},
	\bea
	\textrm{H}(\mathbf{K}_\textrm{M}+\mathbf{k})&=&
	\hbar v_\textrm{F}
	\begin{pmatrix}
		\textbf{k}\cdot \mathbf{\sigma} & 0\\
		0 &  -\textbf{k}\cdot \mathbf{\sigma}^*\\
	\end{pmatrix}, 
	\nonumber
	\\
	\textrm{H}(\mathbf{K}_\textrm{M}'+\mathbf{k})&=&
	\hbar v_\textrm{F}
	\begin{pmatrix}
		- \textbf{k}\cdot \mathbf{\sigma}^* &0\\
		0&  \textbf{k}\cdot \mathbf{\sigma}\\
	\end{pmatrix}, 
	\label{eq:valley}
	\eea
	where $\sigma$ represents the rotated Pauli matrix for the sublattice degree of the freedom. $\mathbf{K}_\textrm{M}$ ( $\mathbf{K}'_\textrm{M}$) represents $\mathbf{K}$($\mathbf{K}'$) point in the moir\'{e} BZ). In general, the addition of arbitrarily small perturbation on the off-diagonal term in Eq. \eqref{eq:valley} gaps out the Dirac magnons. The topological protection of the Dirac magnons requires the decoupling between the particle-hole sector ($U(1)_c$ symmetry),  the decoupling between different valleys ($U(1)_v$ symmetry), and C\textsubscript{2z}$ \mathcal{T}$ (effective space-time inversion), where $\mathcal{T}$ is the effective time-reversal symmetry of magnons ($\mathcal{T}: S_y \rightarrow -S_y $). Within the spin-space group, the coexistence of the three symmetries ($\{ U(1)_c,U(1)_v, C_{2z} \mathcal{T} \}$) provides the $\mathbb{Z}_2$-valued topological protection of the Dirac magnons, associated with the quantized $\pi$-Berry phase.
	
	As the finite interlayer exchange interaction is turned on, Fig. \ref{fig:magnon} (a) shows that the two distinct regions of the collinear FM phase depending on the presence of the magnon gap: gapless small angle regime ($\theta \! \lesssim \!10^\circ$) and gapped large angle regime ($\theta \! \gtrsim \!10^\circ$). The angle dependent magnon gap can be understood by effective decouplings between different valleys \cite{PhysRevX.8.031089}. In the small angle limit, the size of the moir\'{e} unit cell diverges, and the long-wave length component of the interlayer coupling (small momentum exchange) become dominant.
	The finite intervalley couplings, which requires short-wave length coupling (large momentum exchange), are suppressed in the small angle limit. The approximate restoration of $U(1)_v$ symmetry with C\textsubscript{2z} symmetry protects the Dirac magnons.
	
	On the other hands, in generic large angles, the finite valley coupling breaks $U(1)_v$ symmetry.  Fig. \ref{fig:magnon} (a) shows that the non-negligible magnon gap is enhanced as the twist angle is increased ($\theta \rightarrow 30^\circ$). The gapped Dirac magnon is characterized by the non-trivial C\textsubscript{2y}-symmetry protected $\mathbb{Z}$-class topological number, which is given as\cite{PhysRevB.88.075142},
	\bea
	N_\mathbb{Z}=\prod_{n\in occ, C_{2y}=1} \textrm{sgn}(n_\Gamma-n_{M})|n_\Gamma-n_{M}|
	\eea
	where $n_\Gamma$ and $n_{M}$ represents the number of the occupied bands in $C_{2y}=1$ space below the massive Dirac points at $\Gamma$ and C\textsubscript{2y}-symmetric ${M}$ point respectively. In the presence of the additional $C_{2x}$ symmetry which the magnonic Hamiltonian approximately preserves, we can additionally define the Zak phase, $\nu_\pm$, calculated along C\textsubscript{2y} symmetric line for $C_{2y}=\pm 1$ subsector. The mirror winding number  $\nu_\pm$ can be evaluated by ,
	\bea
	\nu_\pm = \frac{1}{i\pi}\log\det[\mathcal{U}_\pm(\Gamma,M_3)\mathcal{U}_\pm(M_3,\Gamma)].
	\eea
	Here,   $\mathcal{U}_\pm(\vv{k}_1,\vv{k}_2) \equiv  P \exp[i\int_{\vv{k}_1}^{\vv{k}_2}\boldsymbol {\mathcal{A}_\pm}(\vv{k})\cdot d\vv{k} ]
	= \tilde{P}_\pm(\vv{k}_1)[\prod_\vv{k} \tilde{P}_\pm(\vv{k})] \tilde{P}_\pm(\vv{k}_2)$, $\mathcal{A}_\pm$ is the non-Abelian Berry connection evaluated from the $\pm$-sector, respectively. $P$ indicates the path ordering, and $\tilde{P}_\pm$ is the projection operator to the occupied  mirror $\pm$-subspace. The physical manifestation of the non-trivial Zak phase is the quantized polarization along C\textsubscript{2y}-symmetric line in the BZ, analogous to the Su-Schrieffer-Heeger model The quantized polarization manifests as the topological magnon modes at the C\textsubscript{2y}-symmetric corners, characterizing the HOTMI phase [See Fig. \ref{fig:magnoncorner}].\cite{PhysRevLett.123.216803,PARK2021260}
	
	\begin{figure}
		\includegraphics[width=0.49\textwidth]{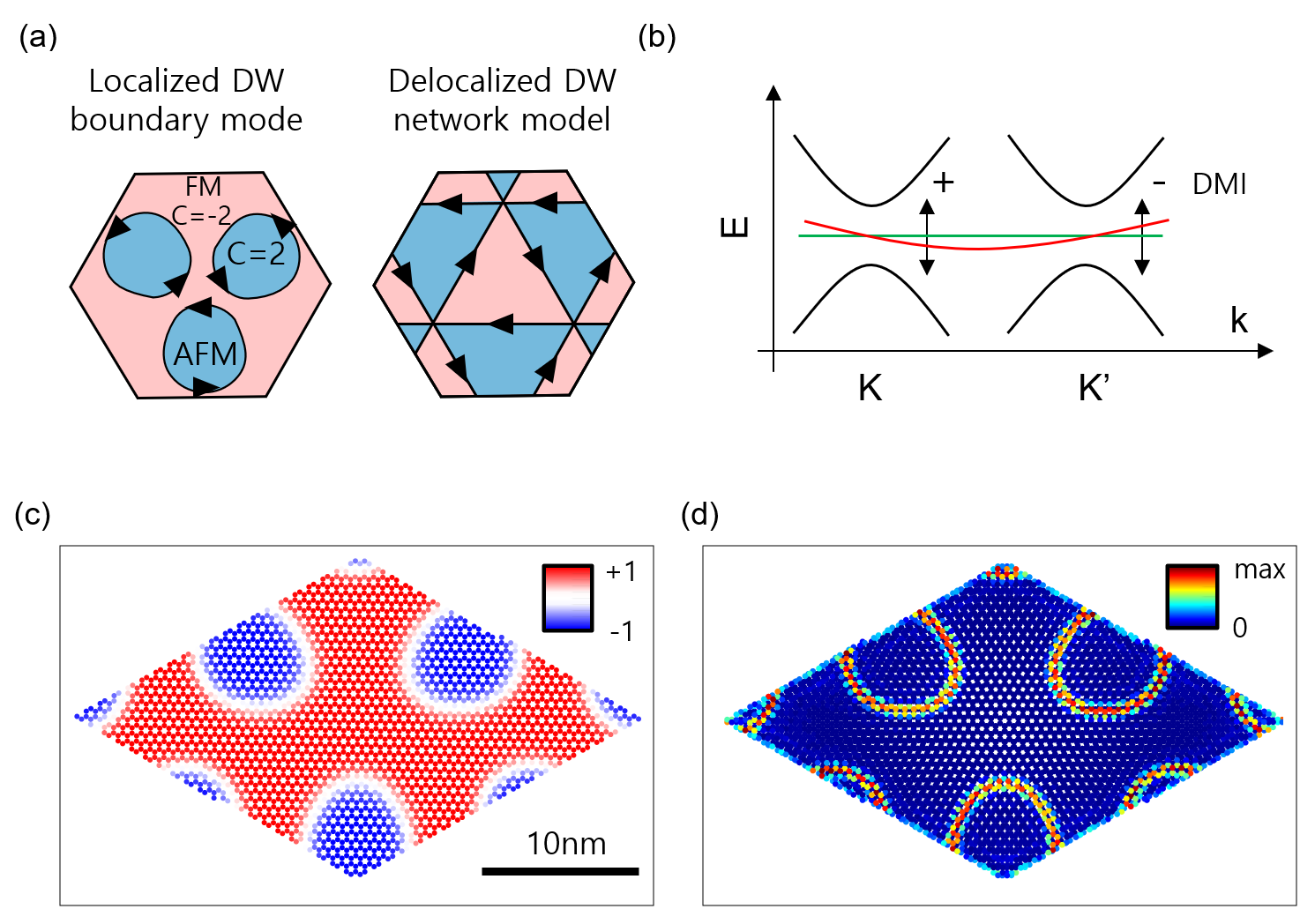}
		\caption{ (a) In the MD phase, the finite DMI induces the different Chern number between the interior and the exterior of the domain. The physical manifestation is the topological edge mode confined in the domain wall. (b) When the domain wall form a closed (open) loop within a unit cell, the magnon spectrum shows the flat (dispersive) band spectrum. (c) the spin texture of the MD phase at the twist angle, $\theta=1.61^\circ$ [(m,n)=(20,21)]. (d) Numerically calculated wave function amplitude of the topological edge mode.  }
		\label{fig:magnonedge}
	\end{figure}

	\subsection{Non-collinear phase : Topological domain wall edge mode}
	
	In the NCD phase, the spin tilting to the azimuthal directions explicitly breaks $U(1)_c$ symmetry, and it contributes to the finite pairing gap in the BdG Hamiltonian of the magnons. The amplitude of the magnon gap is gradually enhanced as the spin tilting to the horizontal direction is increased in the deep NCD phase region. As the twist angle is further lowered, in the MD phase, the local spins of deep inside and far outside of the domain can be approximated as the collinear interlayer antiferromagnet and ferromagnet respectively, where $\pi$-spin flip locally restores $U(1)_c$ symmetry both inside and outside of the magnetic domain. Due to the restoration of $U(1)_c$ symmetry, the magnon gap vanishes.
	
	On the other hand, CrI$_3$ still possesses the additional inversion symmetry preserving next-neareast neighbor intralayer DMI \cite{PhysRevX.8.041028}, which gaps out the Dirac magnons. The DMI is explicitly written as,
	\bea
	H_{\textrm{DMI}}=\sum_{\langle\langle i,j \rangle \rangle} \mathbf{A}_{ij}\cdot \mathbf{S}_i \times \mathbf{S}_j,
	\eea
	where $\mathbf{A}_{ij}=\pm|D|\nu_{ij}\hat{z}$ represent the DMI, where the sign alternates for the sublattice degree of freedom. $\nu_{ij}=+1(-1)$ for the clockwise (counter-clockwise) circulation respect to the honeycomb center. The addition of the small DMI does not suppress the local FM order, but it immediately generates the mass gap of Dirac magnons by breaking the time-reversal symmetry (of the magnonic Hamiltonian). The magnonic Hamiltonian of each layer corresponds to the Haldane model of the electronic system, which is characterized by the non-trivial Chern number\cite{Owerre_2016}. The Chern number changes the sign depending on the relative orientations of the DM vector and the magnetic order. (i.e. $C=1(-1)$ if the DM vector and the spin polarization is parallel (anti-parallel)). As a result, the inside and outside the domain wall has the Chern number difference $+2$, which physically manifests as the pair of the topological edge mode flowing around the domain wall [See Fig. \ref{fig:magnonedge} (d)].
	
	In the moir\'e BZ, the topological edge mode realizes as the extensive number of the flat band (mid gap states within the massive Dirac fermions), since each domain wall magnon is isolated from one another. On the other hands, near the disordered phase by gradually decreasing the intralayer coupling strength, the spin configurations of the domains are connected each other, and the domain wall form extended network (See Fig. \ref{figs:spin_conf_basic} in the Appendix \ref{Appsec:MC}). In such a case, the localized topological edge modes form a delocalized network \cite{PhysRevLett.124.137002}, and the corresponding low energy theory is described by Chalker-Coddington model \cite{Chalker_1988} [Fig. \ref{fig:magnonedge} (b)]. In the moir\'e BZ, the delocalized network states manifests dispersive band.

	
	\section{Conclusion}\label{sec:Conclusion}
	
	In this work, we have constructed the theory of moir\'e magnetism. Starting from the first-principles calculations, we construct the effective spin model of twisted bilayer CrI$_3$. Our theory based on full lattice model descriptions reveal various non-collinear order and topological magnon phases that have been overlooked in previous studies. In particular, the interlayer coupling, which intrinsically lacks $C_{2z}$ symmetry, plays an important role by stabilizing the magnetic skyrmion phase and HOTMI phase without DMI. We also classify the topological magnon phases, and the topological phases sensitively depend on the underlying symmetry of the magnetic order. The measurement of the change of the magnonic excitations can be the fingerprints of the exotic non-collinear orders, and it can be experimentally detected through the thin-film neutron scattering experiment \cite{Cenker2021}.
	
	Our theory can be readily generalized to the other family of the transition metal trihalides such as CrCl$_3$ and CrBr$_3$. Broader classes of two-dimensional magnetic materials are promising candidates for twisted magnetism. For example, the family of A$_2$TMO$_3$ compounds (A=Na,Li, TM=transition metal), which contains many monoclinic honeycomb antiferromagnets \cite{Takagi2019,PhysRevLett.108.127204,Lee_2012,STROBEL198890,TAKAHASHI20081518}, is expected to exhibit qualitatively different behaviors as it exhibits zigzag intralayer order. It would be also an interesting direction to consider a general form of spin interactions. Transition metal phosphorus trisulfides, TMPS$_3$ \cite{19833919,Kim_2019} compounds exhibits Ising type (FePS$_3$) \cite{doi:10.1021/acs.nanolett.6b03052,JERNBERG1984178,Chandrasekharan1994,PhysRevB.35.7097,PhysRevB.94.214407,Kim_2019}, XY type and XXZ type (NiPS$_3$) \cite{kim20192,PhysRevB.92.224408,Kuo2016,PhysRevLett.120.136402,PhysRevB.98.134414}, and Heisenberg-type antiferromagnetic interactions (MnPS$_3$) \cite{PhysRevB.46.5425,RONNOW2000676,Toyoshima_2009,Kim_20192,PhysRevB.82.100408,Wildes_1994,doi:10.1143/JPSJ.55.4456}. More complicated form of magnetic interactions such as Kitaev interactions \cite{PhysRevLett.102.017205,PhysRevB.101.245126} in the moir\'e systems can be also interesting for future studies.
	
	Another natural generalization of our work is to consider the twisted multiple-layer magnets. For instance, in the case of twisted triple layers, enhanced interlayer coupling is expected to promote the non-collinearity and unconventional magnetic phases. In the case of the twisted double bilayer \cite{cheng2022electrically}, we expect that the top and the bottom layers interacts with each other through the renormalized coupling strength. In such case, the effective spin model realizes the strong coupling regime of the moir\'e superlattice, where the strength of the interlayer coupling exceeds the intralayer coupling strength. The extension to the multiple layer system will be the topic of the future study.

	\section{Acknowledgements}
	M.J.P. is grateful to Yong Baek Kim and SungBin Lee for fruitful and enlightening discussions. K. K. and M.J.P. acknowledge financial support from the Institute for Basic Science in the Republic of Korea through the project IBS-R024-D1. M.J.H. was supported by the National Research Foundation of Korea (NRF) grant funded by the Korea government (MSIT) (No. 2021R1A2C1009303 and No. NRF2018M3D1A1058754). 
	
	Note : During the compeletion of the manuscript, we noticed that the higher-order topological magnonic states are independently proposed in the simple ferromagnetic twisted bilayer ferromagnet model in the presence of $C_{2z}$ symmetry \cite{https://doi.org/10.48550/arxiv.2202.12151}. In our work, we explicitly show that the HOTMI phase is well-defined in the absence of $C_{2z}$ symmetry, which is the case for the ferromagnetic phase of CrI$_3$.
	
	\bibliography{reference}

	\newpage 
	\onecolumngrid 
	\newpage 
	\appendix
	\vspace*{\fill}
	{\Large Appendix}
	\vspace*{\fill}
	
	\maketitle
	
	\tableofcontents
	\pagebreak
	\section{Crystal structure of twisted bilayer CrI\textsubscript{3}} \label{Appsec:lattice}
	
	\subsection{Lattice structure of monolayer CrI$_3$}
	
	\begin{figure}[t]
		\centering
		\includegraphics[width=.7\textwidth]{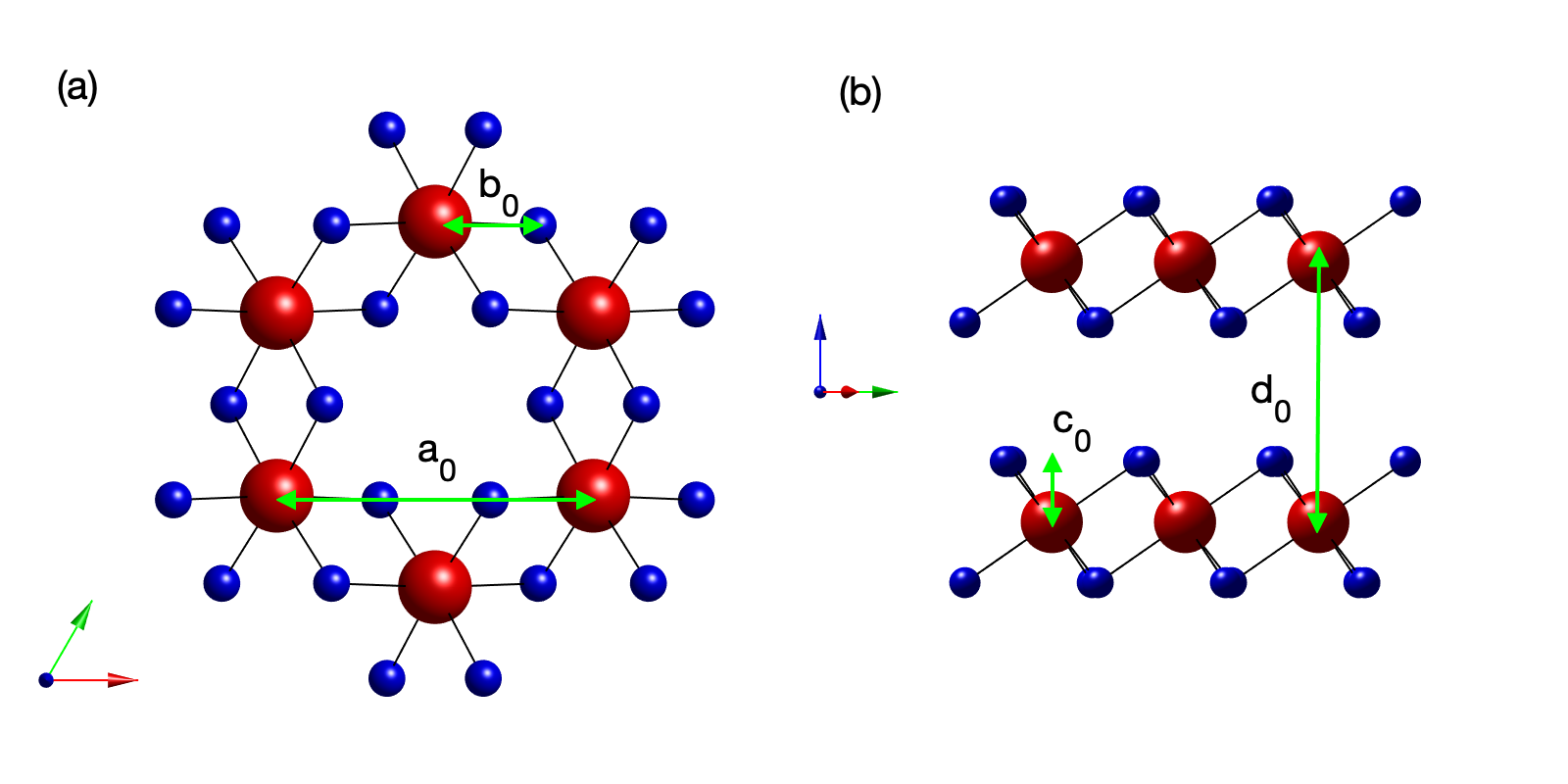}
		\caption{(a) Top view of monolayer CrI\textsubscript{3} and (b) side view of AA-stacked bilayer CrI\textsubscript{3}}
		\label{figs:atomic_struct}
	\end{figure}

	\begin{table}[b!]
		\centering
		\begin{tabular}{c|c|c|c|c}\toprule
			Structure & C\textsubscript{3z} & C\textsubscript{2y} & S\textsubscript{6} & Point Group \\ \midrule
			Monolayer & $+$ & $+$ & $+$ & D\textsubscript{3d} \\
			AA-stacked bilayer & $+$ & $+$ & $+$ & D\textsubscript{3d} \\
			Twisted bilayer with $\mathbf L_\perp$ making hexagonal centers overlapped & $+$ & $+$ & $-$ & D\textsubscript{3} \\
			Twisted bilayer with $\mathbf L_\perp$ making interlayer Cr atoms overlapped & $+$ & $-$ & $-$ & C\textsubscript{3} \\ 
			Twisted bilayer with $\mathbf L_\perp \parallel \hat{x}$ & $+$ & $+$ & $-$ & C\textsubscript{2} \\
			Twisted bilayer with other $\mathbf L_\perp$ & $-$ & $-$ & $-$ &  C\textsubscript{1} \\ \bottomrule
		\end{tabular}
		\caption{Point group symmetries of CrI\textsubscript{3} twisted bilayer structures defined in Eq.~(\ref{Appeq:twisted_bilayer_structure}). ``$+~(-)$" indicates that the corresponding symmetry is present (absent).}
		\label{tab:point_group_crystal}
	\end{table}
	
	Monolayer CrI\textsubscript{3} consists of magnetic Cr atoms that form a honeycomb lattice. Each Cr atom is surrounded by six I atoms that form a distorted edge-sharing octahedron \cite{doi:10.1021/acs.nanolett.8b03321}. The lattice structure is described by the following lattice vector:
	\begin{equation}
		\mathbf R_i = x_i \mathbf a_1 + y_i \mathbf a_2,
	\end{equation}
	where $\mathbf a_1 = a_0(1,0,0)$ and $\mathbf a_2 = a_0 (1/2,\sqrt3/2,0)$ are primitive lattice vectors of a triangular lattice and $x_i$ and $y_i$ are integers and $a_0 = 6.86 \AA$ is the lattice length of the honeycomb lattice. The coordinates of the two Cr atoms in the unit cell are given as,
	\begin{subequations}
		\begin{align}
			\mathbf R_{i,s}^{\textrm{Cr}} & = \mathbf R_i +\mathbf D_s^{\textrm{Cr}}, \\
			\mathbf D_s^{\textrm{Cr}} & = \Big\{ (0,a_0/\sqrt{3},0), ~ (0,0,0)\Big\},
		\end{align}
	\end{subequations}
	where $s = A, B$ denote the sublattice index for the honeycomb lattice of Cr atoms. The coordinates of the additional six I atoms in the unit cell can be expressed as,
	\begin{subequations}
		\label{Eqs:Iodine}
		\begin{align}
			\mathbf R_{i,a}^{\textrm{I},\textrm{l}} & = \mathbf R_i + \mathbf D_{a}^{\textrm{I},\textrm{top(bottom)}}, \\
			\mathbf D_{a}^{\textrm{I},\textrm{top}} & = \Big\{ \big(b_0\cos(60^\circ+\phi_0), b_0\sin(60^\circ+\phi_0), c_0\big),~ \big(b_0\cos(180^\circ+\phi_0), b_0\sin(180^\circ+\phi_0), c_0\big), \nonumber \\  
			& ~~~~~~\big(b_0\cos{(300^\circ+\phi_0)}, b_0\sin{(300^\circ+\phi_0)}, c_0\big)\Big\}, \\
			\mathbf D_{a}^{\textrm{I},\textrm{bottom}} & = \Big\{ \big(b_0\cos(-\phi_0), b_0\sin(-\phi_0), -c_0\big), \big(b_0\cos(120^\circ-\phi_0),b_0 \sin(120^\circ-\phi_0),-c_0\big), \nonumber \\
			& ~~~~~~\big(b_0\cos(240^\circ-\phi_0), b_0\sin(240^\circ-\phi_0), -c_0\big) \Big\},
		\end{align}
	\end{subequations}
	where $b_0=2.24 \AA$ and $c_0=1.56 \AA$ stand for the distance from the center Cr atom to I atom in the in-plane and the z direction respectively, and $\phi_0 = 2.2^\circ $ is a distortion angle.
	
	The corresponding coordinates of monolayer CrI\textsubscript{3} is classified by D\textsubscript{3d} point group, which contains the important symmetries : C\textsubscript{2y}, C\textsubscript{3z} rotational symmetries, and S\textsubscript{6z} roto-reflection symmetry \cite{glazer2012space}. The symmetry operations for each of these symmetries are given as,
	\begin{subequations}
		\begin{align}
			\mathbf R_{i,s}^\textrm{Cr}-\mathbf R_0 & \rightarrow R_\textrm{X} (\mathbf R_{i,s}^\textrm{Cr}-\mathbf R_0), \\
			\mathbf R_{i,a}^{\textrm{I},\textrm{l}}-\mathbf R_0 & \rightarrow R_\textrm{X} (\mathbf R_{i,a}^{\textrm{I},\textrm{l}}-\mathbf R_0),
		\end{align}
	\end{subequations}
	where the transformation matrices $R_\textrm{X}$ for X $\in \{ 
	\textrm{C\textsubscript{2y}}, \textrm{C\textsubscript{3z}}, \textrm{S\textsubscript{6z}}
	\} $ are explicitly written as,
	\begin{equation}
		R_{\textrm{C\textsubscript{2y}}} = \begin{bmatrix} -1 & 0 & 0 \\ 0 & 1 & 0 \\ 0 & 0 & -1\end{bmatrix},~~
		R_{\textrm{C\textsubscript{3z}}} = \begin{bmatrix} -1/2 & \sqrt{3}/2 & 0 \\ -\sqrt{3}/2 & -1/2 & 0 \\ 0 & 0 & 1\end{bmatrix},~~
		R_{\textrm{S\textsubscript{6z}}} = \begin{bmatrix} 1/2 & -\sqrt{3}/2 & 0 \\ \sqrt{3}/2 & 1/2 & 0 \\ 0 & 0 & -1\end{bmatrix}.
	\end{equation}
	Here, $\mathbf R_{0} = (\mathbf a_1 + \mathbf a_2)/3 = a_0 (1/2,\sqrt{3}/6,0)$ is the rotation center located at the hexagonal center. Under such transformations, the positions of Cr and I atoms are invariant up to translation by the lattice vector.

	\subsection{Lattice structure of twisted bilayer CrI$_3$}
	
	In the case of the twisted bilayer, the atomic structure is constructed by the rotation of the optimized AA-stacked structure obtained from Ref. \cite{D1NR02480A}. In the absence of the additional lattice distortion, the twisted bilayer structures of CrI\textsubscript{3} are explicitly constructed as follows,
	\begin{subequations} \label{Appeq:twisted_bilayer_structure}
		\begin{align}
			\mathbf R_{i,s}^{\textrm{Cr},\textrm{t}} & = R (\theta_{m,n}/2)\mathbf R_{i,s}^{\textrm{Cr}} + (d_0/2) \hat{z} + \mathbf L_\perp/2, \\
			\mathbf R_{i,s}^{\textrm{Cr},\textrm{b}} & = R (-\theta_{m,n}/2) \mathbf R_{i,s}^{\textrm{Cr}} - (d_0/2) \hat{z} - \mathbf L_\perp/2, 
			\\
			\mathbf R_{i,a}^{\textrm{I},\textrm{t\textsubscript{1}}} & = R (\theta_{m,n}/2)\mathbf R_{i,a}^{\textrm{I},\textrm{top}} + (d_0/2) \hat{z} + \mathbf L_\perp/2, 
			\\
			\mathbf R_{i,a}^{\textrm{I},\textrm{t\textsubscript{2}}} & = R (\theta_{m,n}/2) \mathbf R_{i,a}^{\textrm{I},\textrm{bottom}} + (d_0/2) \hat{z} + \mathbf L_\perp/2, 
			\\
			\mathbf R_{i,a}^{\textrm{I},\textrm{b\textsubscript{1}}} & = R (-\theta_{m,n}/2) \mathbf R_{i,a}^{\textrm{I},\textrm{top}} - (d_0/2) \hat{z} - \mathbf L_\perp/2, 
			\\
			\mathbf R_{i,a}^{\textrm{I},\textrm{b\textsubscript{2}}} & = R (-\theta_{m,n}/2) \mathbf R_{i,a}^{\textrm{I},\textrm{bottom}} - (d_0/2) \hat{z} - \mathbf L_\perp/2,
		\end{align}
	\end{subequations}
	where the coordinates of the iodine atoms outward the layers $\mathbf R_{i,a}^{\textrm{I},\textrm{t\textsubscript{1}}},\mathbf R_{i,a}^{\textrm{I},\textrm{b\textsubscript{2}}}$ (inward layers $\mathbf R_{i,a}^{\textrm{I},\textrm{t\textsubscript{2}}},\mathbf R_{i,a}^{\textrm{I},\textrm{b\textsubscript{1}}}$) are characterized by the constants in Eq. \eqref{Eqs:Iodine}, $b_0=2.24\AA,c_0=1.56\AA,\phi_0=2.2^\circ$ $(b_0=2.22\AA, c_0=1.57\AA,\phi_0=2.9^\circ)$ respectively. Here, $\theta_{m,n} = \arccos\big[\frac{1}{2}\frac{m^2+n^2+4mn}{m^2+n^2+mn}\big]$ is a commensurate twist angle generated by a pair of the coprime integers $m$ and $n$. $R(\phi)$ is a transformation matrix for rotation about the z axis, which is explicitly given as,
	\begin{equation}
		R(\phi) = \begin{bmatrix} \cos{\phi} & -\sin{\phi} & 0 \\ \sin{\phi} & \cos{\phi} & 0 \\ 0 & 0 & 1 \end{bmatrix}. \label{Appeq:rot_mat}
	\end{equation}
	$d_0 = 6.7 \AA $ is the interlayer distance between the two layers, and $\mathbf L_\perp = (L_x,L_y,0)$ represents a global lateral shift between two layers in the in-plane direction.

	An untwisted AA-stacked bilayer with $\theta_{m,n}=0$ and $\mathbf L_\perp = (0,0,0)$ has the same point group symmetry D\textsubscript{3d} as with the monolayer. A twist by $\theta_{m,n}\neq0$ makes interlayer Cr and I atoms misaligned, so it breaks S\textsubscript{6z} of the original AA-stacked bilayer in addition to the inversion symmetry. As a result, the point group of the twisted bilayer is classified by D\textsubscript{3} point group symmetry or the subgroups of D\textsubscript{3}, which includes C\textsubscript{3}, C\textsubscript{2}, and C\textsubscript{1}, depending on the lateral shift vector $\mathbf L_\perp$. There exist four different cases depending on the lateral shifts: (i) If $\mathbf L_\perp=(0,0,0)$ or the lateral shift by the finite $\mathbf L_\perp$ makes some hexagonal centers of Cr atoms from both layers overlapped on $x-y$ plane, then the twisted bilayer has the D\textsubscript{3} point group symmetry since both C\textsubscript{2y} and C\textsubscript{3z} are preserved. (ii) If $\mathbf L_\perp$ makes two Cr atom sites from the top and the bottom layers overlap on $x-y$ plane, then the twisted bilayer has the C\textsubscript{3} point group symmetry since the new rotation center for C\textsubscript{3z} appears at this position. (iii) If $\mathbf L_\perp \parallel \hat{x}$, the twisted bilayer has the C\textsubscript{2} point group symmetry since C\textsubscript{2y} symmetry is preserved. (iv) Finally, if $\mathbf L_\perp$ does not satisfy any of the conditions above, it breaks both C\textsubscript{2y} and C\textsubscript{3z}, then the twisted bilayer has the trivial point group symmetry of C\textsubscript{1}. Table~\ref{tab:point_group_crystal} summarizes the point group symmetries of all possible twisted bilayer structures and their point group symmetries.

	\section{Spin model construction} \label{Appsec:model}
	
	\subsection{Coupling function for interlayer exchange interactions} \label{Appsec:J_inter}

	In this section, we characterize the form of the interlayer exchange couplings of the twisted bilayer CrI\textsubscript{3} based on the crystalline symmetry. While the local magnetic moments of CrI$_3$ are carried by the magnetic Cr atoms, the non-magnetic I atoms also effect the crystalline symmetries by participating the exchange couplings of the Cr spins through the super-super exchange interactions \cite{doi:10.1021/acs.nanolett.8b03321}. In general, there exist four different interlayer coupling interaction function for the different kinds of the sublattice degree of the freedom $(J_\textrm{AA}^\perp (\mathbf{r}),J_\textrm{AB}^\perp (\mathbf{r}),J_\textrm{BA}^\perp (\mathbf{r}),J_\textrm{BB}^\perp (\mathbf{r}))$ as a function of the relative coordinate displacement of two Cr atoms, $\mathbf r \equiv \mathbf{R}^{\textrm{Cr,top}} - \mathbf{R}^{\textrm{Cr,bottom}}$, where $\mathbf{R}^{\textrm{Cr,top(bottom)}}$ is the coordinate vector of Cr atom in the top (bottom) layer. 
	
	The form of the interlayer exchange coupling function is determined by the underlying point group symmetry of the atomic structure. Firstly, in the non-twisted limit, the inversion symmetry of the atomic structure forces  
	\bea
	J_\textrm{AA}^\perp (\mathbf{r})=J_\textrm{BB}^\perp (\mathbf{r}), \quad J_\textrm{AB}^\perp (\mathbf{r})=J_\textrm{BA}^\perp (\mathbf{r}).
	\eea Therefore, we only need to consider the two independent interlayer exchange couplings. Secondly, the interlayer exchange coupling should be invariant under C\textsubscript{3z} rotation operation as
	\begin{equation}
		J_{ss'}^\perp (\mathbf r) = J_{ss'}^\perp \big(R(2\pi/3)\mathbf r\big) = J_{ss'}^\perp \big(R(4\pi/3)\mathbf r\big), \label{Appeq:J_inter_relation1}
	\end{equation}
	where $R(\phi)$ is given in Eq.~(\ref{Appeq:rot_mat}). Thirdly, the reflection operations about three planes, each of which is parallel with $\mathbf r_m = \big\{(1,0), (1/2,\sqrt{3}/2), (1/2,-\sqrt{3}/2) \big\}$, gives rise to the following condition as
	\begin{equation}
		J_{ss'}^\perp (\mathbf r) = J_{ss'}^\perp \big(M(\mathbf r_m)\mathbf r \big), \label{Appeq:J_inter_relation2}
	\end{equation}
	where $M(\mathbf r) = \frac{1}{|\mathbf r|^2} \begin{bmatrix} r_{x}^2-r_{y}^2 & 2r_{x} r_{y} \\ 2 r_{x} r_{y} & r_{y}^2 - r_{x}^2 \end{bmatrix}$.
	
	We can write down the functional form of the interlayer exchange coupling, which satisfies both Eqs.~(\ref{Appeq:J_inter_relation1}) and (\ref{Appeq:J_inter_relation2}), as
	\begin{equation}
		J_{ss'}^\perp (\mathbf r) = J_0 \exp\big(-|r-d_0|/l_0\big) + J_1^s \exp\big(-|r-r_*|/l_1^s\big) \sum_{a=1}^{3}\sin(\mathbf q_a^s \cdot \mathbf r_{\parallel}) + J_1^c \exp\big(-|r-r_*|/l_1^c\big) \sum_{a=1}^{3}\cos(\mathbf q_a^c \cdot \mathbf r_{\parallel}). \label{Appeq:J_inter}
	\end{equation}
	Here, the first term describes the isotropic decaying coupling term while the second and the third sinusoidal functions describe the oscillatory behavior of the interlayer coupling depending on the variation of the local stacking structure. The two sets of wave vectors $\mathbf q_{a}^{s,c}$ describe the detailed pattern of such change, where the C\textsubscript{3z} symmetry in Eq.~(\ref{Appeq:J_inter_relation1}) forces the wave vectors in each set to be related with each other, while their directions are further specified by the reflection symmetry in Eq.~(\ref{Appeq:J_inter_relation2}). As a result, the wave vectors are given explicitly as,
	\begin{subequations}
		\begin{align}
			& \mathbf q_{1}^{s}  = q^s\big(1,0,0\big),~~~~~~~~~~ \mathbf q_{2}^{s}  = R(2\pi/3) \mathbf q_{1}^{s}, ~~ \mathbf q_{3}^{s} = R(4\pi/3) \mathbf q_{1}^{s}, \\
			& \mathbf q_{1}^{c}  = q^c\big(\sqrt{3}/2,1/2,0\big),~~ \mathbf q_{2}^{c} = R(2\pi/3) \mathbf q_{1}^{c}, ~~ \mathbf q_{3}^{c} = R(4\pi/3) \mathbf q_{1}^{c}.
		\end{align}
	\end{subequations}
	We have introduced the exponential functions to describe the decaying of the interlayer coupling as a function of distance $r = \sqrt{|\mathbf r_\parallel|^2+d_0^2}$, where $d_0$ is the constant interlayer distance. Physically, such decay originates from the decaying of the hybridization of orbitals as a function of interlayer distance in the super-super exchange interaction \cite{doi:10.1021/acs.nanolett.8b03321}. The remaining unspecified parameters including the interaction strength, $J_0, J_1^s, J_1^c$, the magnitudes of the wave vectors, $q^{s}, q^{c}$, and the other length scales, $l_0, l_1^s, l_1^c, d_0, r_*$, are determined by the DFT calculations.
	
	\subsection{Ab initio parameters and DFT calculation details} \label{Appsec:DFT}

	\begin{figure}[t!]
		\includegraphics[width=\textwidth]{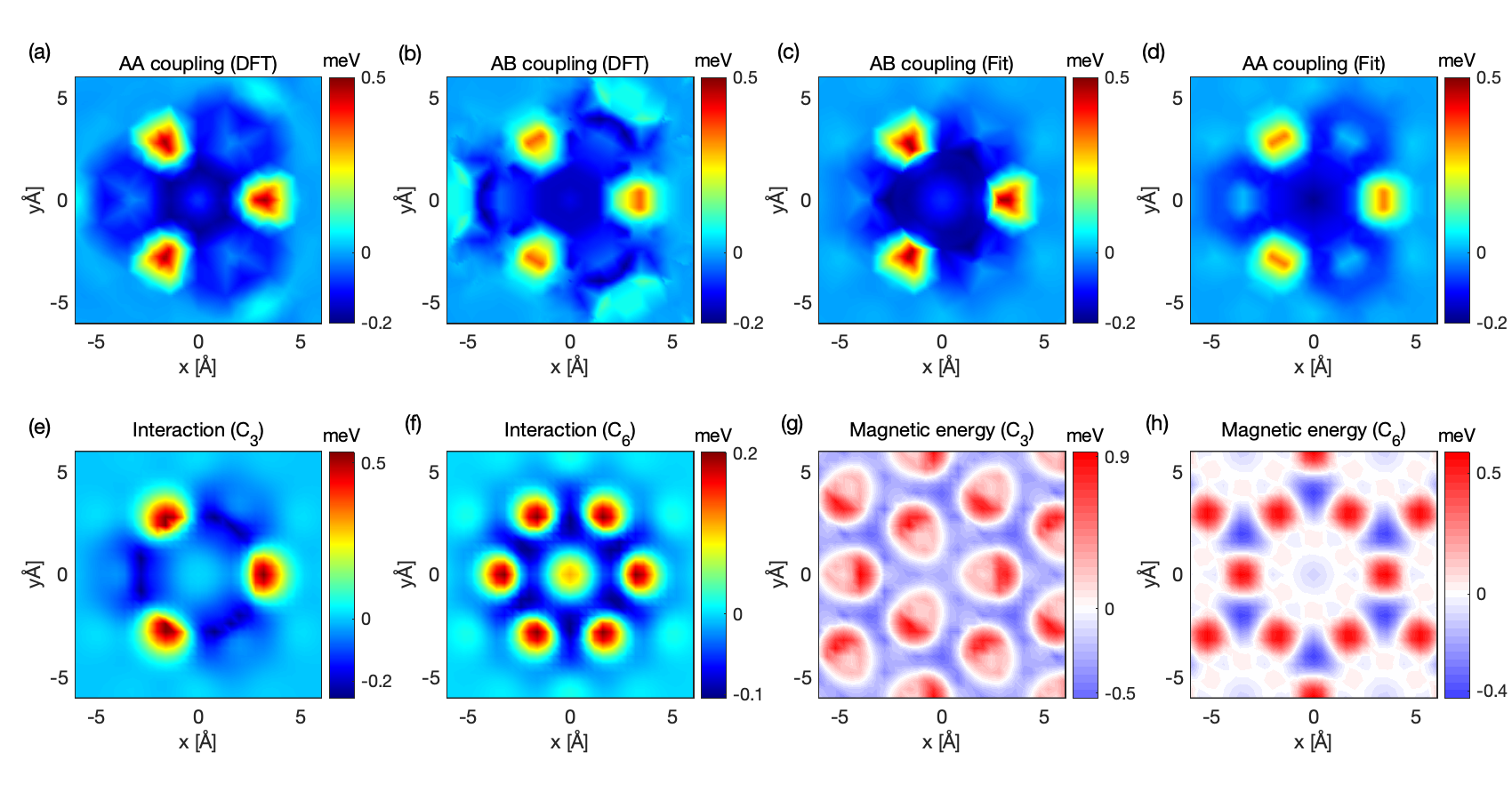}
		\caption{(a)-(d) Interlayer Heisenberg exchange interactions in bilayer CrI\textsubscript3 as a function of perpendicular displacement $(x,y)$ between two interlayer Cr spins. (a) and (b) present DFT results for interactions between spins on the same sublattice and on different sublattices, respectively. (c) and (d) present the fitting results for (a) and (b), respectively, where the fitting function is given in Eq.~(\ref{Appeq:J_inter}) and the fitting parameters are given in the first and second rows in Table~\ref{tabS:J_inter}. (e)-(h) Interlayer Heisenberg exchange interactions in two simple models studied in this work, which are constructed based on the DFT results. (e) and (f) present C\textsubscript{3}- and C\textsubscript{6}- symmetric interlayer interactions, respectively, constructed from Eq.~(\ref{Appeq:J_inter}) with the parameters given in the third and fourth rows in Table~\ref{tabS:J_inter}. (g) and (h) present the stacking dependent the interlayer magnetic energy corresponding to (e) and (f), respectively, where $x, y$ denote the displacement for a global relative translation of two layers in the perpendicular direction. The negative (positive) sign indicates ferromagnetic (antiferromagnetic) interactions in all plots.}
		\label{figs:J_inter}
	\end{figure}
	
	\begin{table}[b!]
		\begin{tabular}{c|cccccccc}
			\toprule
			& $\textrm{J}_0 (\textrm{meV})$ &$\textrm{J}^\textrm{s}_1 ('') $ & $\textrm{J}^\textrm{c}_1 ('') $ & $l_0 (\textrm{\AA})$ & $l^\textrm{s}_1 ('')$ & $l^\textrm{c}_1  ('')$ & $\mathbf q^\textrm{s}_{1} (\textrm{\AA}^{-1})$ & $\mathbf q^\textrm{c}_{1} ('')$ \\ \midrule
			AA coupling (DFT) &  $-0.20$ & $-0.66$ & $0.10$ &  $0.35$ & $0.23$ & $0.59$ & $(0.65,0)$  & $(1.78,1)$  \\
			AB coupling (DFT) & $-0.31$ & $-0.24$ & $0.10$ & $0.12$ & $0.40$ & $0.65$ & $(0.72,0)$ & $(1.82,1)$ \\ \midrule
			C\textsubscript{3}-symmetric (model) & $-0.1$ & $-0.5$ & $0.1$ & $0.1$ & $0.3$ & $0.6$ & $(0.7,0)$ & $(1.73,1)$ \\
			C\textsubscript{6}-symmetric (model) & $0$ & $0$ & $0.1$ & - & - & $0.6$ & - & $(1.73,1)$ \\
			\bottomrule
		\end{tabular}
		\caption{The parameters for the coupling function in Eq.~(\ref{Appeq:J_inter}) for interlayer Heisenberg interactions. The first and second rows present the fitting results of DFT data. The third and fourth rows present the parameter values for two simplified models studied in this work. }
		\label{tabS:J_inter}
	\end{table}
	
	We now determine the precise form of Eq.~(\ref{Appeq:J_inter}) based on the density-functional theory computations. We used the density functional theory package ‘openMX’ \cite{PhysRevB.69.195113,
		PhysRevB.67.155108}, which takes the linear combination of numerical pseudo-atomic orbital (PAO) as a basis set and the norm-conserving pseudo-potential. DFT calculations were performed within the generalized gradient approximation \cite{PhysRevLett.78.1396} for the exchange-correlation potential. To describe the on-site electron correlation within Cr 3d orbitals, the DFT +U method was adopted with a charge-density-based scheme \cite{PhysRevB.93.045133,PhysRevB.91.241111,PhysRevB.92.035146,Ryee2018,Ryee_2018} using U = 2.9 eV and J = 0.7 eV from cRPA method \cite{PhysRevMaterials.3.031001}. The 11×11×1 k-meshes are used for the calculation. The vacuum size is taken by larger than 18.6 $\AA$. The structure is constructed by the lateral shift of the AA-stacked structure \cite{D1NR02480A}. The magnetic exchange interactions between intra- and interlayer Cr ions were calculated within the magnetic force linear response calculation \cite{Oguchi_1983,PhysRevB.54.1019,LIECHTENSTEIN198765,PhysRevB.61.8906,PhysRevB.70.184421} using the implemented package ‘Jx’ \cite{PhysRevB.97.125132,YOON2020106927}
	
	We obtain the real space map of the interlayer Heisenberg exchange coupling as a function of perpendicular displacement between two Cr atoms as shown in Fig.~\ref{figs:J_inter} (a) and (b). We fit the DFT data to the coupling function in Eq.~(\ref{Appeq:J_inter}) and obtain the fitting parameters as summarized in Table~\ref{tabS:J_inter} in the first and second rows. Fig.~\ref{figs:J_inter} (c) and (d) present the fitting results for (a) and (b), respectively, which agree well with the DFT results.

	Based on the above analysis, we construct two simplified models studied in this work. The first model, which is the main consideration in this work, retains the C\textsubscript{3}-symmetric feature of the original interlayer coupling of bilayer CrI\textsubscript{3} [see Fig.~\ref{figs:J_inter} (e)] and exhibits the similar pattern of stacking-dependent interlayer magnetism [see Fig.~\ref{figs:J_inter} (g)] as with bilayer CrI\textsubscript{3} \cite{doi:10.1021/acs.nanolett.8b03321}. The second model exhibits the C\textsubscript{6}-symmetric feature [see Fig.~\ref{figs:J_inter} (g)], and accordingly the C\textsubscript{6}-symmetric pattern of stacking dependent interlayer magnetism [see Fig.~\ref{figs:J_inter} (h)]. In both models, the same interlayer coupling function in Eq.~(\ref{Appeq:J_inter}) is utilized where the parameters are given in the third and fourth rows in Table~\ref{tabS:J_inter}, respectively. In both models, we do not distinguish AA and AB couplings since the difference between them is minor in bilayer CrI\textsubscript{3} as shown in Fig.~\ref{figs:J_inter} (a) and (b).
	
	\section{Monte-Carlo simulation} \label{Appsec:MC}
	
	\subsection{Simulation method} \label{Appsec:simulation_method}
	
	\begin{figure}[t]
		\centering
		\includegraphics[width=\textwidth]{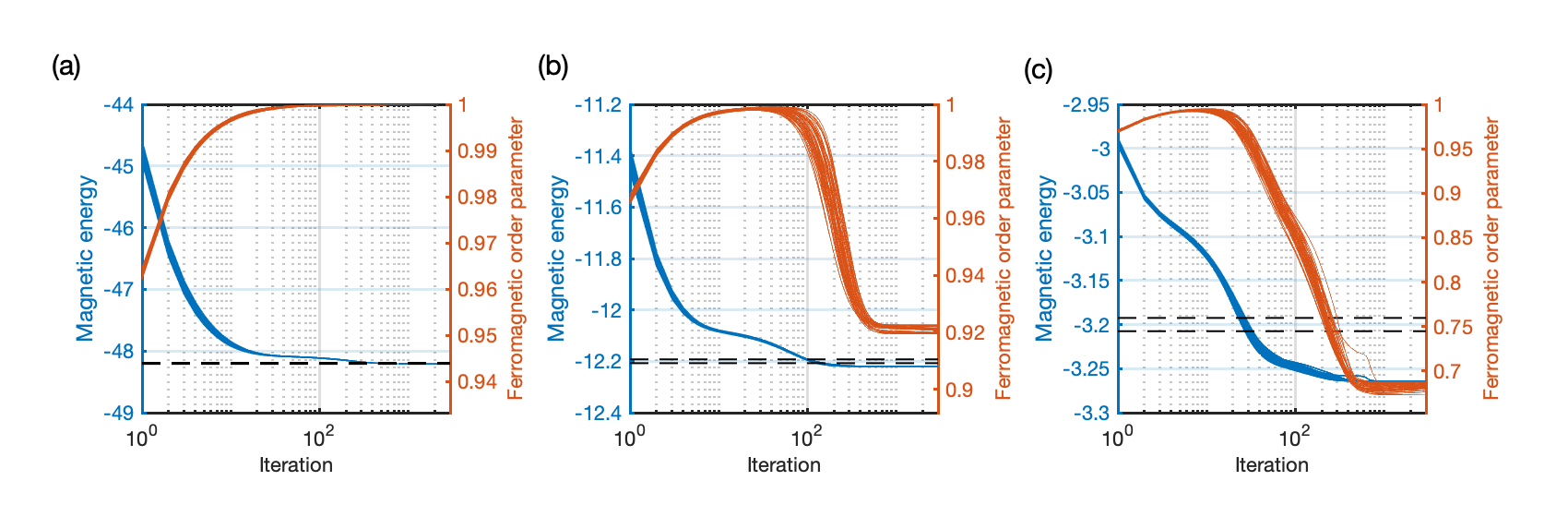}
		\caption{Convergence tests of the Monte-Carlo simulation in three magnetic phases: a collinear ferromagnetic phase (a), a non-collinear domain phase (b), and a magnetic domain phase (c). $J = 16,~2,~1$ are used for (a), (b) and (c), respectively. $\theta = 2.0^\circ$ and $D_z = 0.2$ are utilized in all plots. Each lines represent the different sets of the initial conditions.}
		\label{figs:MC_convergence}
	\end{figure}
	
	To find the magnetic ground state of the classical spin model, we perform an extensive Monte-Carlo approach. Here, we focus on translation-invariant magnetic states satisfying periodic boundary conditions, since there is no next neighbor couplings in the unit of the moir\'e unit cell, which induces the magnetic order with finite $\mathbf{q}$ vector. For the detailed method, we first construct a random spin configuration from $N$ spin vectors $\mathbf n_i$, each of which is randomly generated. Next, we update the spin configuration by subtracting each $\mathbf n_i$ with a gradient vector $h_i^a \equiv \frac{\partial H}{\partial n_i^a}$ and normalizing it as
	\begin{equation}
		\mathbf n_i \rightarrow \mathbf n_i' = \frac{\mathbf n_i - c \mathbf h_i }{|\mathbf n_i - c \mathbf h_i|},
	\end{equation}
	where $c$ is a control parameter for the update rate and the gradient term $\mathbf h_i$ is explicitly given as
	\begin{subequations}
		\begin{align}
			h_i^x & = - 2J \sum_{j\in \mathcal N(i)}n_j^x + 2\sum_j J_{ij}^\perp n_j^x, \\
			h_i^y & = - 2J \sum_{j\in \mathcal N(i)}n_j^y + 2\sum_j J_{ij}^\perp n_j^y, \\
			h_i^z & = - 2J \sum_{j\in \mathcal N(i)}n_j^z + 2\sum_j J_{ij}^\perp n_j^z - 2D_z n_i^z.
		\end{align}
	\end{subequations}
	The new spin configuration with $\mathbf n_i'$ has a lower energy than the original one with $\mathbf n_i$. After a sufficiently large number of iterative updates, the resulting configuration converges and becomes a local minimum of the energy functional. To find a global minimum one, we take a large number of samples, each of which has a different random initial configuration, and perform the same computation for each sample. Then, the ground state of the spin model is chosen by the minimum energy one among the resulting minimum energy configurations. The result is stochastically correct and becomes more accurate as more samples are taken into account. 
	
	Figure \ref{figs:MC_convergence} shows the convergence tests of the Monte-Carlo simulation method. After 5000 sweeps of iterations, the magnetic state converges to (a) a collinear ferromagnetic state, (b) a non-collinear domain state, and (c) a magnetic domain state where the convergence is manifested by both the magnetic energy per spin $H[\mathbf n_i]/N$ (left axis) and the ferromanetic order parameter $\left| \sum_{i}\mathbf n_i\right|/N$ (right axis). Here, we have taken 50 samples most of which converge to the same magnetic state in fact.
	
	\subsection{Ground state spin configurations} \label{Appsec:spin_conf}
	
	\begin{figure}[t]
		\centering
		\includegraphics[width=\textwidth]{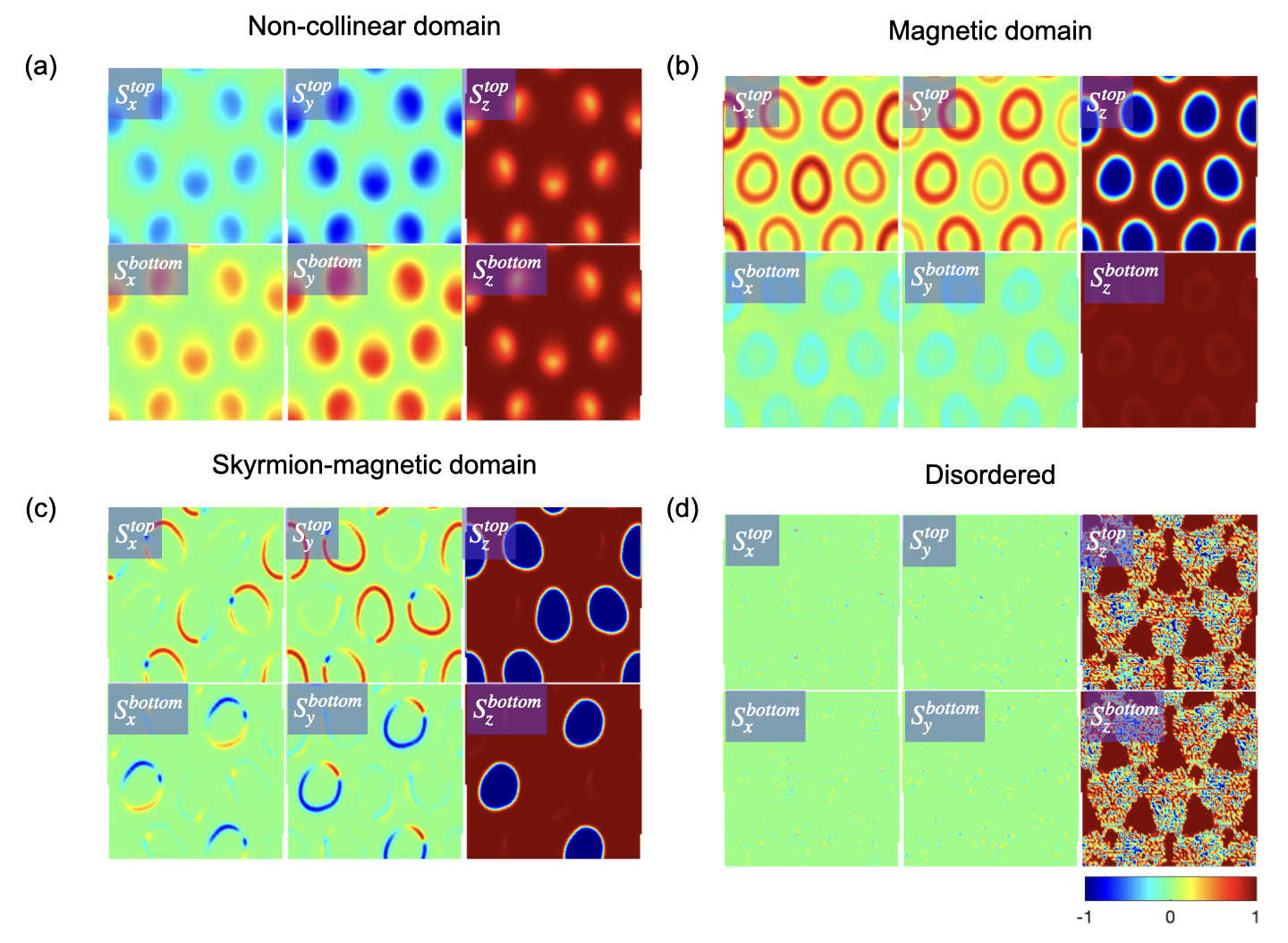}
		\caption{Detailed spin configurations in (a) the non-collinear domain (NCD), (b) magnetic domain (MD), (c) skyrmion-magnetic domain (Sk-MD), and (d) disordered phases, respectively. $J = 16, 4, 1, 0.03 $ meV are used for (a)-(d), respectively. $\theta = 0.93^\circ$ and $D_z=0.2$ meV are used in all plots. }
		\label{figs:spin_conf_basic}
	\end{figure}
	
	Figure \ref{figs:spin_conf_basic} shows the ground state spin configurations presented in Fig.~\ref{fig:MC} and \ref{fig:skyrmion} in more details. In (a) NCD phase, the spin configuration exhibits the interlayer antiferromagnetism in the in-plane direction. In the interior of the domain, the spins in two layers are tilted in the opposite direction while in the exterior of the domain they are ordered ferromagnetically in the out-of-plane direction. Meanwhile, in (b) MD phase, the spin configuration exhibits the interlayer antiferromagnetism in the out-of-plane direction. In the interior (exterior) of the domain, the spins in two layers are antiferromagnetically (ferromagnetically) ordered in the out-of-plane direction. Accordingly, in the spin-flipped layer (the top layer in the example) the spins rotate 180{\textdegree} across the domain while in the other layer the spins are almost fixed, showing only the slight tilting within the domain wall. In (c) Sk-MD phase, the spin configuration exhibits the similar antiferromagnetism as MD phase. The difference is that the spins in the domain walls rotate horizontally and wind through the equator in the Bloch sphere. The overall spin texture exhibits a spiral skyrmion configuration, which distinguishes this phase from MD phase with the plain domain wall. Finally, in (d) disordered phase, the interlayer spins are aligned parallel or anti-parallel in an almost random fashion depending on the sign of the local interlayer coupling. 
	
	\begin{figure}[t]
		\centering
		\includegraphics[width=.9\textwidth]{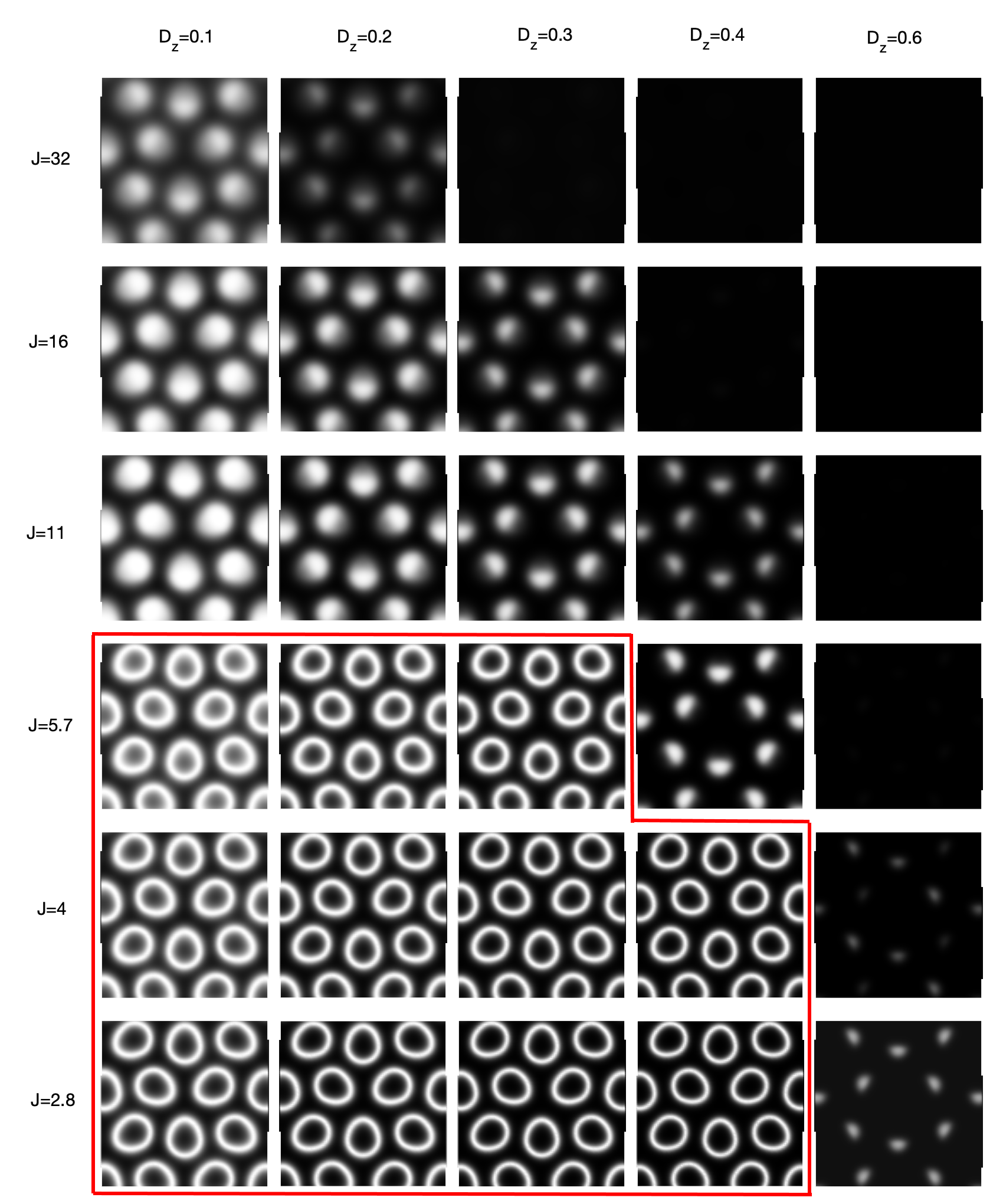}
		\caption{Examples of ground state spin configurations in NCD phase (outside the red box) and MD phase (inside the red box) for selected $J$ and $D_z$. Each panel represents one of two layers (the spin-flipped layer) in NCD phase (MD phase). In each panel, the black-while color code indicates the magnitude of the out-of-plane component of the local spin where the black implies that the spin is polarized in the z direction. $\theta = 0.93^\circ$ is utilized in all plots. }
		\label{figs:spin_conf_evol}
	\end{figure}
	
	Figure \ref{figs:spin_conf_evol} shows selected examples of the ground spin configurations from which we computed the order parameters presented in Fig.~\ref{fig:MC}, \ref{fig:global_phase_diagram}, and \ref{fig:ContinuumModel} of the main text. A series of panels shows the evolution of the characteristic features of magnetic states such as the tilting angle, the domain size, and the domain-wall width as a function of $J$ and $D_z$. In NCD phase (outside the red box), the tilting angles of spins increase as $J$ decreases as shown by the increasing of the brightness in the domain. Meanwhile, the size of the domain continuously decreases as $D_z$ increases as shown by the decreasing size of the white disk. In MD phase (inside the red box), the size of the domain (the size of the ring) is not very dependent on $J$ nor $D_z$. Meanwhile, the width of the domain wall (the width of the white ring) does depend on $J$ and $D_z$, and it continuously decreases as $J$ decreases or $D_z$ increases.

	\subsection{Order parameters of phases transitions} \label{Appsec:order_params}
	
	\begin{figure}[t]
		\centering
		\includegraphics[width=\textwidth]{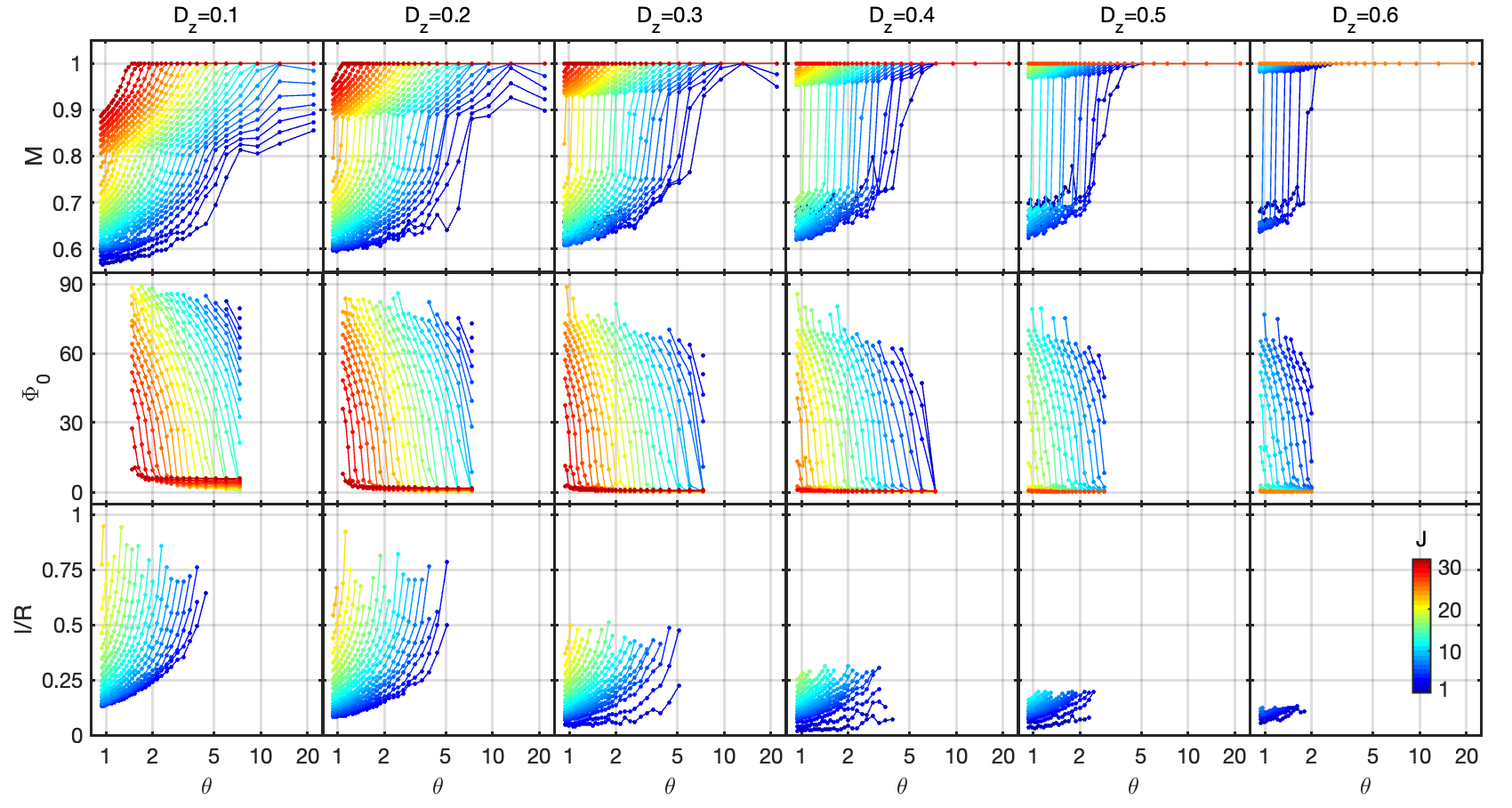}
		\caption{ Three order parameters computed from the Monte-Carlo simulation. Here, $M$ in the first row stands for the ferromagnetic order parameter defined in Eq.~(\ref{Appeq:M_def}), $\Phi_0$ in the second row stands for the tilting order parameter in NCD phase defined in Eq.~(\ref{Appeq:Phi_def}), and $l/R$ stands for the domain-wall width order parameter in MD phase defined in Eq.~(\ref{Appeq:l/R_def}). The evolution of the order parameters is given as a function of $\theta$. The color scale represents $J$ value for each curve where the scale is shown in the color bar at the right-bottom plot. }
		\label{figs:order_params}
	\end{figure}
	
	The observations made in Sec.~\ref{Appsec:spin_conf} suggest that the NCD state can be characterized by the tilting angle while the MD state can be characterized by the width of the domain wall. Those quantities change systematically as a function of $J$ and $D_z$ so they can be utilized as ``order parameters" for the magnetic phase transitions. In addition to the usual ferromagnetic order parameter, we compute those three order parameters to investigate the magnetic phase transitions quantitatively. The ferromagnetic order parameter is computed numerically as
	\begin{equation} 
		M = \frac{1}{N}\left| \sum_{i} \mathbf n_i \right|, \label{Appeq:M_def}
	\end{equation}
	where $\mathbf n_i$ is a spin vector at site $i$ and $N$ is the number of spins in the moir\'e unit cell. The tilting angle order parameter is computed numerically as
	\begin{equation} 
		\Phi_0 = \textrm{max} \big\{ \Phi_{i} \big\}, \label{Appeq:Phi_def}
	\end{equation}
	where $\Phi_{i} = \cos^{-1} \Big[n_i^z/\sqrt{(n_i^x)^2+(n_i^y)^2}\Big]$ is the polar angle of $\mathbf n_i$. The order parameter for the domain-wall width is computed numerically as
	\begin{equation} 
		l/R = 1-\sqrt{1-N_{dw}/(N_d+N_{dw})}, \label{Appeq:l/R_def}
	\end{equation}
	where $l$ and $R$ are the width of the domain-wall and the radius of the whole domain, respectively. Here, the order parameter is defined as the ratio of $l$ and $R$ since both $l$ and $R$ increase trivially as the twist angle $\theta$ decreases but their ratio is not scaled with $\theta$ so can utilized to keep track of the evolution of MD state. If the shape of the domain is approximated as a disk [see Fig.\ref{figs:spin_conf_basic} (b)], $l$ and $R$ are given as $2\pi Rl-\pi l^2 = A_0 N_{dw}$ and $\pi (R-l)^2 = A_0 N_{d}$ where $N_{dw}$ and $N_{d}$ are the number of spins inside the domain wall and the interior of the domain, respectively, and $A_0$ is a unit area. From the relations, Eq.~(\ref{Appeq:l/R_def}) can be obtained straightforwardly. We numerically find $N_{dw}$ by counting the number of tilted spins whose out-of-plane components are smaller than some threshold $\lambda_{dw}$, i.e. $|n_i^z|< \lambda_{dw}$ ($|n_i^z|\geq \lambda_{dw}$). We also find $N_{d}$ by counting the number of flipped spins, i.e. $n_i^z * (\sum_i n_i ^z) < 0$, which are not in the domain wall, i.e. $|n_i^z| \geq \lambda_{dw}$. We take $\lambda_{dw}$ to be $0.9$ and the result is not very dependent on the value of $\lambda_{dw}$.
	
	Figure \ref{figs:order_params} shows the computation results of the order parameters. As shown in the first row, the ferromagnetic parameter well describes global magnetic phase transitions. The continuous FM-NCD phase transition is signified by decreasing of $M$, which is fixed to be unity in FM phase. The discontinuous NCD-MD phase transition is signified by discontinuous drop of $M$. The size of drop decreases as $D_z$ decreases and finally turns into a kink as $D_z \rightarrow 0$, indicating that NCD and MD phases coalesce and become indiscernible. As shown in the second row in Fig.~\ref{figs:order_params}, the tilting angle order parameter $\Phi_0$ correctly describes the FM-NCD phase transition, which is zero in FM phase and increases continuously in NCD phase. Finally, as shown in the third row in Fig.~\ref{figs:order_params}, the domain wall width order parameter $l/R$ correctly describes the evolution of MD state where $l/R$ continuously decreases as $\theta$ decreases in MD phase.

	\section{Continuum model} \label{Appsec:CM}
	
	
	In this section, we derive the free energy expressions given in Eq.~(\ref{eq:F_NCD}) and Eq.~(\ref{eq:F_MD}) of the main text. Consider the moir\'e magnets with small twist angle. In such system, the local spin vector is found to vary smoothly accordingly with the smooth change of the interlayer coupling inside the moir\'e unit cell, and the gradient of the local spin vector can be considered as a small perturbation. In this case, we may take the continuum limit approximation of the classical spin model and expand it with the gradient of the local spin vector. As a result, we obtain the following free energy functional:
	\begin{align}
		F[\mathbf n_t, \mathbf n_b] & = \frac{1}{A_0}\int d \mathbf x \sum_{l=t,b} \bigg\{\frac{3a_0^2J}{2}[\nabla \mathbf n_l (\mathbf x) ]^2 - D_z [n_l^z (\mathbf x)]^2\bigg\} +\frac{1}{A_0}\int d \mathbf x  ~ J_\perp(\mathbf x) \mathbf n_t(\mathbf x) \cdot \mathbf n_b(\mathbf x), \label{Appeq:free_energy_functional}
	\end{align}
	where $\mathbf n_{t,b}(\mathbf x) = \big(\sin\theta_{t,b}(\mathbf x) \cos \phi_{t,b}(\mathbf x),\sin\theta_{t,b}(\mathbf x) \sin\phi_{t,b}(\mathbf x),\cos \theta_{t,b}(\mathbf x)\big)$ represents a normalized vector field for spins in each layer. $a_0$ and $A_0=3\sqrt{3}a_0^2/4$ are the lattice length and the unit area of the monolayer of honeycomb lattice, respectively. The integral is constrained on the magnetic domain region where spin configuration changes across magnetic phase transitions.
	
	\subsection{Free energy of non-collinear domain (NCD) phase} \label{Appsec:F_NCD}
	
	We first compute the free energy $F$ for the non-collinear domain phase. As observed in the simulation, this phase is characterized by the formation of local non-collinear magnetic domains [see Fig.~\ref{figs:spin_conf_basic} (a)]. Suppose spins are aligned in the positive z-direction, $\mathbf n_{t,b} = (0,0,1)$, outside these domains. Then, inside each domain, the spins are progressively tilted from the positive $z$-direction to the horizontal direction, where the orientation of the tilting is opposite in two layers. For an analytic computation, we take into account this configuration in a simple function form given as
	\begin{subequations} \label{Appeq:spin_conf_NCD}
		\begin{align}
			\mathbf n_t & = \big(\sin[\Phi_0(1-\rho/R)],0,\cos[\Phi_0(1-\rho/R)]\big), \\
			\mathbf n_b & = \big(-\sin[\Phi_0(1-\rho/R)],0,\cos[\Phi_0(1-\rho/R)]\big),
		\end{align}
	\end{subequations}
	where $\Phi_0$ stands for the maximum value of the tilting angles, $R$ is the width of the non-collinear magnetic domain, and $\rho$ is a radial coordinate measured from the center of the domain. $\phi_t=0$ and $\phi_b=-\pi$ have been taken without loss of generality ($\phi_t - \phi_b = \pi $ in general). Inserting Eq.~(\ref{Appeq:spin_conf_NCD}) into Eq.~(\ref{Appeq:free_energy_functional}), we obtain
	\begin{align}
		F(\Phi_0) & = \frac{1}{A_0}\int^{2\pi}_{0}d\varphi\int^{R}_{0}d\rho \rho\bigg\{2\tilde{J}(\Phi_0/R)^2 - 2D_z \cos^2[\Phi_0(1-\rho/R)] + J_\perp(\rho,\varphi) \cos [2\Phi_0(1-\rho/R)]\bigg\}.\label{Appeq:F_NCD_integral}
	\end{align}
	The integrals for $\tilde{J}$ and $D_z$ are computed straightforwardly as
	\begin{equation}
		\int^{2\pi}_{0}d\varphi\int^{R}_{0}d\rho \rho\bigg\{2\tilde{J}(\Phi_0/R)^2-2D_z \cos^2[\Phi_0(1-\rho/R)]\bigg\}=2\pi\tilde{J}\Phi_0^2-\pi R^2D_z(1+\sin^2\Phi_0/\Phi_0^2). \label{Appeq:F_NCD_integral1}
	\end{equation}
	The integral for $J_\perp(\rho,\varphi)$ depends on the form of $J_\perp(\rho,\varphi)$ and cannot be evaluated in general. However, when the spin vector fields vary smoothly, i.e. $\Phi_0/R\ll1$, we may obtain the generic functional dependence on $J_\perp(\rho,\varphi)$ by approximating the integral as
	\begin{equation}
		\int^{2\pi}_{0}d\varphi \int^{R}_0 d\rho \rho J_\perp(\rho,\varphi) \cos[2\Phi_0(1-\rho/R)] \simeq 2\pi \bar J_\perp\int^{R}_0 d\rho \rho \cos[2\Phi_0(1-\rho/R)] = \pi R^2 \Big( \bar J_\perp \sin^{2}\Phi_0/\Phi_0^2\Big), \label{Appeq:F_NCD_integral2}
	\end{equation}
	where $\bar J_\perp$ is the average value of $J_\perp(\rho,\varphi)$ given as
	\begin{equation}
		\bar J_\perp \equiv \frac{1}{\pi R^2}\int^{2\pi}_{0}d\varphi \int^{R}_0 d\rho \rho J_\perp(\rho,\varphi). \label{Appeq:J_perp_av}
	\end{equation}
	Inserting Eqs.~(\ref{Appeq:F_NCD_integral1}) and (\ref{Appeq:F_NCD_integral2}) into Eq.~(\ref{Appeq:F_NCD_integral}), we finally obtain
	\begin{equation}
		F(\Phi_0) = \frac{4\pi}{\sqrt3}J\Phi_0^2 -N_{ncd}(\theta) D_z (1+\sin^{2}\Phi_0/\Phi_0^2) + N_{ncd}(\theta) \bar J_\perp \sin^{2}\Phi_0/\Phi_0^2,
	\end{equation}
	where $N_{ncd}(\theta)$ is the number of tilted spins in the non-collinear magnetic domain, which is given as $N_{ncd}(\theta) = \pi R^2/A_0$. 
	
	The free energy $F(\Phi_0)$ explains the FM-NCD phase transition within a standard Landau theory of the phase transition. As $\Phi_0$ increases, $F$ is increased by the first two terms while decreased by the last term. When $F$ is dominated by the last term, which can be achieved by either decreasing $J$ and/or $D_z$ or increasing $N_{ncd}$, a continuous phase transition occurs and $F$ has a non-trivial minimum with a non-zero $\Phi_0$. Near the phase transition, $\Phi_0$ would be small, and we may expand $F$ about $\Phi_0=0$ as
	\begin{equation}
		F(\Phi_0) = F_0 + \Phi_0^2 \bigg\{\frac{4\pi}{\sqrt3}J-\frac{N_{ncd}(\theta)}{3}(-D_z+\bar J_\perp)\bigg\} + \Phi_0^4 \bigg\{ \frac{2N_{ncd}(\theta)}{45}(-D_z + \bar J_\perp )\bigg\} +\mathcal O(\Phi_0^6),
	\end{equation}
	where $F_0$ is a free energy in the ferromagnetic phase given as
	\begin{equation}
		F_0 = N_{ncd}(\theta)(-2D_z + \bar J_\perp).
	\end{equation}
	We rewrite $F$ more simply as
	\begin{equation}
		F(\Phi_0) = F_0 + \frac{a}{2}\big[J-J_c(\theta)\big] \Phi_0^2 + \frac{b}{4}J_c(\theta) \Phi_0^4 + \mathcal O(\Phi_0^6),
	\end{equation}
	where $J_c(\theta)$ is given by
	\begin{equation}
		J_c(\theta) = c N_{ncd}(\theta)(-D_z + \bar J_\perp),
	\end{equation}
	and $a,b,c$ are given by
	\begin{equation}
		a = \frac{8\pi}{\sqrt{3}}, ~ b = \frac{32\pi}{15\sqrt{3}}, ~c = \frac{1}{4\pi\sqrt{3}}.
	\end{equation}
	We find the ground state by minimizing $F$ with respect to $\Phi_0$ as
	\begin{equation}
		F_{min} = F_0 - \frac{a^2}{4b}J_c(\theta)\big[1-J/J_c(\theta)\big]^2,
	\end{equation}
	where $\Phi_0$ is given by
	\begin{equation}
		\Phi_0 = \pm \sqrt{\frac{a}{b}\big[1-J/J_c(\theta)\big]}.    
	\end{equation}
	Note that the NCD phase terminates at $\bar J_\perp = D_z$. Physically, this means that there is no energy gain from the interlayer coupling in forming non-collinear order since there is a comparable energy cost from $D_z$. 
	
	Lastly, we compute the ferromagnetic order parameter defined in Eq.~(\ref{Appeq:M_def}), which in the continuum model reads as
	\begin{subequations} \label{Appeq:M_def_CM}
		\begin{align}
			\mathbf M & = \mathbf M_1 + \mathbf M_2, \\
			\mathbf M_1 & = \frac{1}{N}\sum_{l=t,b}\frac{1}{A_0}\int_{AF} d\mathbf x ~\mathbf n_l(\mathbf x), \label{Appeq:M1_def} \\
			\mathbf M_2 & = \frac{1}{N}\sum_{l=t,b}\frac{1}{A_0}\int_{F} d\mathbf x ~\mathbf n_l(\mathbf x), \label{Appeq:M2_def}
		\end{align}
	\end{subequations}
	where $\mathbf M_1$ and $\mathbf M_2$ represent the contributions from the magnetic domains and the outside of the magnetic domains, respectively, and $N$ stands for the total number of spins in both layers per moir\'e unit cell. Inserting Eq.~(\ref{Appeq:spin_conf_NCD}) into Eq.~(\ref{Appeq:M1_def}), we obtain
	\begin{equation}
		M_{1,z} = 3\times \frac{2}{NA_0}\int_0^{2\pi} d\varphi \int^R _0 d\rho \rho \cos [\Phi_0(1-\rho/R)] = f_{ncd}\frac{\sin^2(\Phi_0/2)}{(\Phi_0/2)^2},
	\end{equation}
	where $M_{1,x}$ and $M_{1,y}$ are zero. Here, the factors of $3$ come from the fact that there are three magnetic domains in a moir\'e unit cell and $f_{ncd} = 6 N_{ncd}/N$ stands for the fraction of the spins in the domains among all spins in the moir\'e unit cell. Outside the magnetic domains, the spin vector field is given by $\mathbf n_{t,b} = (0,0,1)$. Inserting this into Eq.~(\ref{Appeq:M2_def}), we obtain
	\begin{equation}
		\mathbf M_2 = (0,0,1-f_{ncd}),
	\end{equation}
	Summing both contributions, we finally obtain $M = |\mathbf M|$ as
	\begin{equation}
		M = 1-f_{ncd} + f_{ncd}\frac{\sin^2(\Phi_0/2)}{(\Phi_0/2)^2}.
	\end{equation}
	The ferromagnetic order parameter for the non-collinear domains is given as
	\begin{equation}
		M_{ncd} \equiv (1-M)/f_{ncd} = \frac{\sin^2(\Phi_0/2)}{(\Phi_0/2)^2}.
	\end{equation}

	\subsection{Free energy of magnetic domain (MD) phase} \label{Appsec:F_MD}
	
	We next compute the free energy $F$ for the magnetic domain phase. As observed in the simulation, this phase is characterized by the formation of local antiferromagtic domains where interlayer spins are ordered antiferromagnetically in the out-of-plane direction [see Fig.~\ref{figs:spin_conf_basic} (b)]. Suppose spins are aligned in the positive z direction, $\mathbf n_{t,b}=(0,0,1)$, outside these domains. Then, inside each domain, spins are flipped in one of the layers, say the top layer to be specific, as
	\begin{subequations} \label{Appeq:spin_conf_MD1}
		\begin{align}
			\mathbf n_t(0\leq \rho < R-l) & = (0,0,-1), \\
			\mathbf n_b(0\leq \rho < R-l) & = (0,0,1),
		\end{align}
	\end{subequations}
	where $R$ and $l$ are the width of the domain and the domain-wall, respectively, and $\rho$ is a radial coordinate measured from the center of the domain. In the domain-wall, the top layer spins are progressively flipped while the bottom layer spins are tilted in the opposite direction with the top layer spins. For an analytic computation, we take into account these configurations in simple function forms given as
	\begin{subequations} \label{Appeq:spin_conf_MD2}
		\begin{align}
			\mathbf n_t(R-l < \rho \leq R) & = \big( \sin[\pi(R-\rho)/l],0,\cos[\pi(R-\rho)/l]\big), \\ 
			\mathbf n_b(R-l < \rho \leq R) & = \big( -\sin[\Phi_0(l/2-|\rho-R+l/2|)/l],0,\cos[\Phi_0(l/2-|\rho-R+l/2|)/l]\big),
		\end{align}
	\end{subequations}
	where $\Phi_0$ stands for the maximum value of the tilting angles of the bottom layer spins and $\phi_t=0$ and $\phi_b=\pi$ have taken without loss of generality ($\phi_t-\phi_b = \pi$ in general). Inserting Eqs.~(\ref{Appeq:spin_conf_MD1}) and (\ref{Appeq:spin_conf_MD2}) into Eq.~(\ref{Appeq:free_energy_functional}), we obtain
	\begin{align} \label{Appeq:F_MD_integral}
		F(l, \Phi_0) & = \frac{1}{A_0}\int^{2\pi}_{0}d\varphi\int^{R-l}_{0}d\rho \rho\Big\{-2D_z-J_\perp(\rho,\varphi)\Big\}+\frac{1}{A_0}\int^{2\pi}_{0}d\varphi\int^{R}_{R-l}d\rho \rho\Big\{\tilde{J}(\pi^2+\Phi_0^2)/l^2-D_z \cos^2[\pi(R-\rho)/l]\Big\} \nonumber \\
		& +\frac{1}{A_0}\int^{2\pi}_{0}d\varphi\int^{R-l/2}_{R-l}d\rho \rho\Big\{-D_z \cos^2[\Phi_0(R-l-\rho)/l] +J_\perp(\rho,\varphi) \cos [(\pi-\Phi_0)(R-\rho)/l+\Phi_0]\Big\} \nonumber \\
		& +\frac{1}{A_0}\int^{2\pi}_{0}d\varphi\int^{R}_{R-l/2}d\rho \rho \Big\{-D_z \cos^2[\Phi_0(R-\rho)/l]+J_\perp(\rho,\varphi) \cos [(\pi+\Phi_0)(R-\rho)/l]\Big\}.
	\end{align}
	The integrals for $D_z$ are computed straightforwardly as
	\begin{subequations} \label{Appeq:F_MD_integral1}
		\begin{align}
			\int^{2\pi}_{0}d\varphi\int^{R-l/2}_{R-l}d\rho \rho\Big\{-D_z \cos^2[\pi(R-\rho)/l]\Big\}
			& = -\frac{1}{2}D_z (2\pi Rl-\pi l^2), \\
			\int^{2\pi}_{0}d\varphi\int^{R-l/2}_{R-l}d\rho \rho\Big\{-D_z \cos^2[\Phi_0(R-l-\rho)/l]\Big\} & = -\frac{1}{4}D_z (2\pi Rl-\pi l^2)\bigg( 1+\frac{\sin \Phi_0}{\Phi_0}\bigg) + \frac{1}{4} D_z \pi l^2 \frac{1-\cos\Phi_0}{\Phi_0^2}, \\
			\int^{2\pi}_{0}d\varphi\int^{R}_{R-l/2}d\rho \rho \Big\{-D_z \cos^2[\Phi_0(R-\rho)/l]\Big\} & = -\frac{1}{4}D_z (2\pi Rl-\pi l^2)\bigg(1+\frac{\sin \Phi_0}{\Phi_0}\bigg) - \frac{1}{4} D_z \pi l^2 \frac{1-\cos\Phi_0}{\Phi_0^2}.
		\end{align}
	\end{subequations}
	We again approximate the integrals for $J_\perp(\rho,\varphi)$ to obtain the generic functional dependences on $J_\perp(\rho,\varphi)$ as
	\begin{subequations} \label{Appeq:F_MD_integral2}
		\begin{align}
			& \int^{2\pi}_{0}d\varphi \int^{R-l}_{0} d\rho \rho ~J_\perp(\rho,\varphi) \simeq  \bar J_\perp \pi(R-l)^2, \\
			& \int^{2\pi}_{0}d\varphi\int^{R-l/2}_{R-l}d\rho \rho~J_\perp(\rho,\varphi) \cos [(\pi-\Phi_0)(R-\rho)/l+\Phi_0] \simeq \bar J_\perp 2\pi l^2 \frac{1-\sin[\Phi_0/2]}{(\pi-\Phi_0)^2}-\bar J_\perp(2\pi Rl-\pi l^2)\frac{\cos[\Phi_0/2]}{(\pi-\Phi_0)}, \\
			& \int^{2\pi}_{0}d\varphi\int^{R}_{R-l/2}d\rho \rho J_\perp(\rho,\varphi) \cos [(\pi+\Phi_0)(R-\rho)/l] \simeq \bar J_\perp 2\pi l^2 \frac{1+\sin[\Phi_0/2]}{(\pi+\Phi_0)^2}+\bar J_\perp (2\pi Rl-\pi l^2) \frac{\cos[\Phi_0/2]}{(\pi+\Phi_0)},
		\end{align}
	\end{subequations}
	where $\bar J_\perp$ is the average value of $J_\perp(\rho,\varphi)$ given in Eq.~(\ref{Appeq:J_perp_av}). Inserting Eqs.~(\ref{Appeq:F_MD_integral1}) and (\ref{Appeq:F_MD_integral2}) into Eq.~(\ref{Appeq:F_MD_integral}), we obtain
	\begin{equation}
		F(l,\Phi_0) = K + V, \label{Appeq:F_MD}
	\end{equation}
	where $K$ and $V$ are given by
	\begin{subequations}
		\begin{align}
			K & = (2R/l-1)\frac{2\pi}{\sqrt{3}} J(\pi^2+\Phi_0^2) , \\
			V & = - N_{md}(\theta) (1-l/R)^2 (\bar J_\perp + 2D_z), \nonumber \\
			& + N_{md}(\theta)\Big[2l/R-(l/R)^2\Big] \Bigg\{-D_z\bigg(1+\frac{\sin \Phi_0}{2\Phi_0}\bigg)+\bar J_\perp\bigg(\frac{\cos [\Phi_0/2]}{\pi+\Phi_0}-\frac{\cos [\Phi_0/2]}{\pi-\Phi_0} \bigg) \Bigg\} \nonumber \\
			& + N_{md}(\theta) (l/R)^2 \bar J_\perp \Bigg\{ \frac{2\big(1+\sin[\Phi_0/2]\big)}{(\pi+\Phi_0)^2}+\frac{2\big(1-\sin[\Phi_0/2]\big)}{(\pi-\Phi_0)^2} \Bigg\},
		\end{align}
	\end{subequations}
	where $N_{md}(\theta)$ represents the number of spins in the domain, which is given as $N_{md}(\theta) = \pi R^2 /A_0$.

	Lastly, we compute another order parameter for the NCD-MD phase transition, which is the ferromagnetic order parameter defined in Eq.~(\ref{Appeq:M_def}). Inserting Eqs.~(\ref{Appeq:spin_conf_MD1}) and (\ref{Appeq:spin_conf_MD2}) into Eq.~(\ref{Appeq:M1_def}), we find
	\begin{subequations}
		\begin{align}
			M_{1,x} & = \frac{3}{NA_0}\int^{2\pi}_{0}d\varphi\int^{R}_{R-l}d\rho \rho\Big\{\sin\big[\pi(R-\rho)/l\big]-\sin\big[\Phi_0(l/2-|\rho-R+l/2|)/l\big]\Big\}, \\
			M_{1,z} & = \frac{3}{NA_0}\int^{2\pi}_{0}d\varphi\int^{R}_{R-l}d\rho \rho\Big\{\cos\big[\pi(R-\rho)/l\big]+\cos\big[\Phi_0(l/2-|\rho-R+l/2|)/l\big]\Big\},
		\end{align}
	\end{subequations}
	where $M_{2,y}$ is zero. The integrals can be done straightforwardly, and the results are given as
	\begin{subequations} \label{Appeq:M1_MD}
		\begin{align} 
			M_{1,x} & = \frac{f_{md}}{2} \Big[2l/R-(l/R)^2\Big]\bigg\{ \frac{2}{\pi}-\frac{2}{\Phi_0}(1-\cos[\Phi_0/2])\bigg\}, \\
			M_{1,z} & = \frac{f_{md}}{2} (l/R)^2\frac{4}{\pi^2} + \frac{f_{md}}{2} \Big[2l/R-(l/R)^2\Big] \frac{2\sin[\Phi_0/2]}{\Phi_0},
		\end{align}
	\end{subequations}
	where $f_{md} = 6 N_{md}/N$ stands for the fraction of the spins in the domains among all spins in the moir\'e unit cell. $\mathbf M_2$ is the same with NCD phase, which is given as
	\begin{equation} \label{Appeq:M2_MD}
		\mathbf M_2 = (0,0,1-f_{md}),
	\end{equation}
	Summing both contributions, we finally obtain $\mathbf M$ as
	\begin{subequations}
		\begin{align}
			M_x & = \frac{f_{md}}{2}\Big[2l/R-(l/R)^2\Big] \bigg\{\frac{2}{\pi}-\frac{2}{\Phi_0}(1-\cos[\Phi_0/2])\bigg\},\\
			M_z & = (1-f_{md}) + \frac{f_{md}}{2}(l/R)^2\frac{4}{\pi^2} +  \frac{f_{md}}{2} \Big[2l/R-(l/R)^2\Big]\frac{2\sin[\Phi_0/2]}{\Phi_0},
		\end{align}
	\end{subequations}
	where $M_y$ is zero. Finally, we obtain the ferromagnetic order parameter $M = |\mathbf M|$ as
	\begin{align}
		M & = \Bigg[ \bigg\{\frac{f_{md}}{2}\Big[2l/R-(l/R)^2\Big] \bigg( \frac{2}{\pi}-\frac{2}{\Phi_0}(1-\cos[\Phi_0/2])\bigg) \bigg\}^2 \nonumber \\
		& + \bigg\{(1-f_{md}) + \frac{f_{md}}{2}(l/R)^2\frac{4}{\pi^2} + \frac{f_{md}}{2}\Big[2l/R-(l/R)^2\Big]\frac{2\sin[\Phi_0/2]}{\Phi_0}\bigg\}^2\Bigg]^{1/2}. \label{Appeq:M_MD}
	\end{align}
	
	\subsection{Scaling analysis for the phase transitions} \label{Appsec:scaling_analysis}
	
	\begin{figure}[t!]
		\centering
		\includegraphics[width=\textwidth]{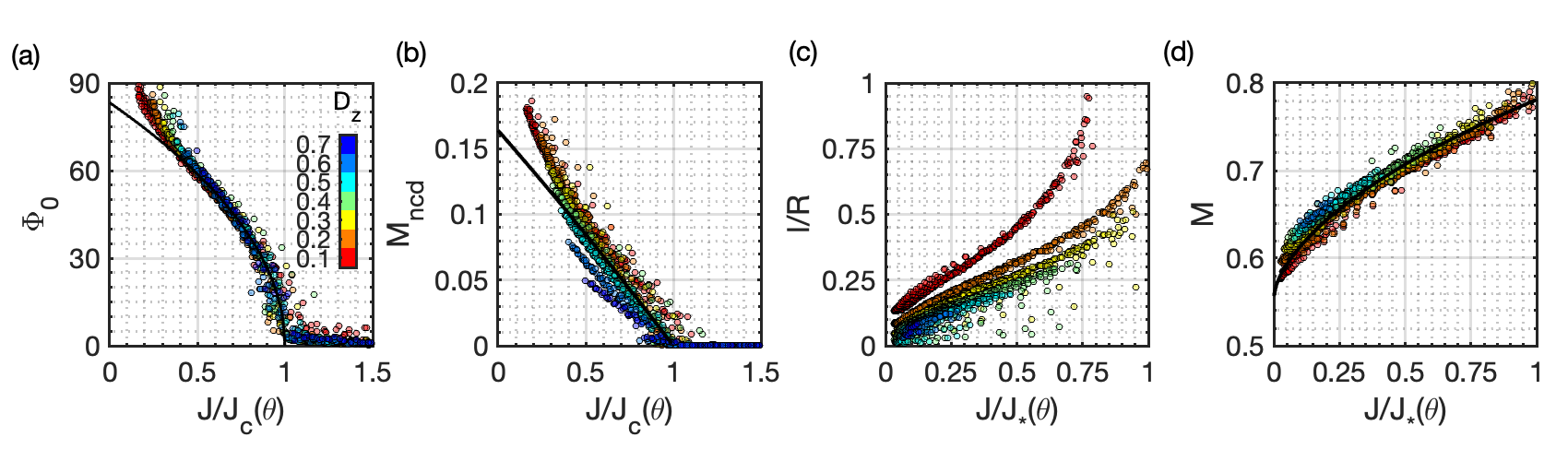}
		\caption{Scaling analysis in the FM-NCD phase transition with respect to (a) the tingle angle order parameter $\Phi_0$ and (b) the ferromagnetic order parameter $M_{ncd}$ in the non-collinear domain. Here, $J_c(\theta)$ represents the critical intralayer coupling strength given in Eq.~(\ref{Appeq:Jc_FM-NCD}) and the black lines represent the analytic fitting functions given in Eqs.~(\ref{Appeq:Phi_scaling}) and (\ref{Appeq:M_ncd_scaling}), respectively. Scaling analysis in the NCD-MD phase transition for (c) the domain-wall width order parameter $l/R$ and (d) the ferromagnetic order parameter $M$. Here, $J_*(\theta)$ represents another critical intralayer coupling strength given in Eq.~(\ref{Appeq:Jc_NCD-MD}). }
		\label{figs:scaling_analysis}
	\end{figure}
	
	Figure \ref{figs:scaling_analysis} (a) and (b) shows the scaling analysis results for FM-NCD phase transition with two order parameters $\Phi_0$ and $M_{ncd}$ [(a) is the same with Fig.~ \ref{fig:ContinuumModel} (a) of the main text]. The scaling formulae for $\Phi_0$ and $M_{ncd}$ are given as
	\begin{subequations} \label{Appeq:FM-NCD_scaling}
		\begin{align}
			\Phi_0 & = \sqrt{(a/b)[1-J/J_c(\theta)]}, \label{Appeq:Phi_scaling} \\
			M_{ncd} & = \frac{\sin^2(\Phi_0/2)}{(\Phi_0/2)^2},\label{Appeq:M_ncd_scaling} \\
			J_c(\theta) & = c f_{ncd} N(\theta) (-D_z+\bar J_\perp), \label{Appeq:Jc_FM-NCD}
		\end{align}
	\end{subequations}
	where $N(\theta) \simeq 4/\theta^2$ is the total number of spins inside the moiré unit cell (valid when $\theta\ll1$), and $a,b,c$ numerical constants [see Sec.~\ref{Appsec:F_NCD} for the derivation]. As shown in the phase diagram of Fig.~\ref{fig:global_phase_diagram} of the main text, the NCD phase terminates at $D_z \approx 0.8$, which suggests $\bar J_\perp \approx 0.8$ since the termination occurs at $\bar J_\perp = D_z $ in the continuum model. The remaining parameter $f_{ncd}$ stands for the fraction of spins in the non-collinear magnetic domain and turns out to strongly vary depending on $D_z$ in the simulation [see Fig.~\ref{figs:spin_conf_evol}]. Thus, we utilize it as a fitting parameter in the scaling analysis for each $D_z$ and find its value as shown in Table~\ref{tab:f_ncd}. As shown in Fig. \ref{figs:scaling_analysis} (a) and (b), the scattered data from different $\theta$ presented in the first and second rows in Fig.~\ref{figs:order_params} are well-collapsed into a single curve which also agrees well with the analytic formula.  

	Figure \ref{figs:scaling_analysis} (c) and (d) shows the scaling analysis results for $l/R$ and $M$ in MD phase. In contrast to the FM-NCD case, the scaling formulae for $l/R$ and $M$ are very difficult to find due to a complex function form of the free energy in Eq.~(\ref{Appeq:F_MD}). Despite the complexity, we conjecture that the competing energy scale for the intralayer coupling $J$ is simply proportional to $\bar J_\perp$ and the number of spins in the magnetic domain $N_{md}(\theta)$, which is explicitly given as 
	\begin{equation}
		J_*(\theta) = \frac{\sqrt{3}}{4\pi^3}f_{md}N(\theta)\bar J_\perp,  \label{Appeq:Jc_NCD-MD}
	\end{equation}
	where $f_{md} \approx 0.55$ stands for the fraction of spins in the magnetic domain and the numerical constant is found from Eq.~(\ref{Appeq:F_MD}). Indeed, as shown in Fig. \ref{figs:scaling_analysis} (c) and (d), we find that the scattered data from different $\theta$ presented in the first and third rows in Fig.~\ref{figs:order_params} are well-collapsed though in different curves for each $D_z$ in contrast to the FM-NCD transition case.
	
	In Fig.~\ref{fig:ContinuumModel} (d) of the main text, we used $J_c(\theta)$ for both NCD and MD phases and $M$ (not $M_{ncd}$) to cover all magnetic phases. The collapse of scattered data from different $\theta$ is clearly seen though in different curves for each $D_z$.

	\begin{table}[b!]
		\centering
		\begin{tabular}{c|c|c|c|c|c|c|c}\toprule
			$D_z$  & 0.1 & 0.2 & 0.3 & 0.4 & 0.5 & 0.6 & 0.7 \\ \midrule
			$f_{ncd}$ & 0.99 & 0.68 & 0.53 & 0.38 & 0.30 & 0.23 & 0.20 \\ \bottomrule
		\end{tabular}
		\caption{The fraction of spins in the non-collinear magnetic domain as a function $D_z$ found from the scaling analysis in Fig.~\ref{figs:scaling_analysis} in (a) and (b).}
		\label{tab:f_ncd}
	\end{table}

	\section{Magnon excitations}\label{Appsec:magnon1}
	\subsection{Derivation of Magnon Hamiltonian}
	
	For a given magnetic ground state, the Bogoliubov-de Gennes (BdG) Hamiltonian of the magnons can be derived by using the Holstein-Primakoff (HP) transformation \cite{PhysRev.58.1098}.
	In this section, we explain how we obtain magnon excitations from the spin Hamiltonian in Eq.~(\ref{eq:spinH}) for a given magnetic order. We first explain our analytic derivation for magnon. We generally write down the Heisenberg spin Hamiltonian with the anisotropy term, which is given as,
	\begin{subequations}
		\begin{align}
			H & = \sum_{i,j}\sum_{\mu,\nu}H_{ij\mu\nu}+\sum_{i}\sum_{\mu}A_{i\mu}, \\
			H_{ij\mu\nu} & = \sum_{\alpha=x,y,z}J_{ij\mu\nu}S_{i\mu}^\alpha S_{j\nu}^\alpha,\\
			A_{i\mu}&= -D_z S_{i\mu}^zS_{i\mu}^z.
		\end{align}
	\end{subequations}
	where $H_{ij\mu\nu}$ and $A_{i\mu}$ represent a Heisenberg interaction and a single-ion anisotropy energy. $i,j$ is the index represents the unit cell. $\mu,\nu$ represent the different sites within a unit cell. $\alpha=x,y,z$ is the index that represent the direction of the magnetic order. We consider a generic non-collinear magnetic order where the spin vectors of $\mathbf n_{\mu}$ are ordered differently at each site $\mu$. To perform the HP transformation, we need to transform the spin operators at each site represented in the global coordinate to the local coordinate in the ordered direction by rotating the coordinate system. The spin operator, $\tilde{S}$, represented in the local coordinate frame, is given as,
	\begin{equation}
		S_{i\mu}^\alpha = \sum_{\beta=x,y,z}[R(\mathbf n_{\mu})]^{\alpha\beta}\tilde{S}_{i\mu}^\beta,
	\end{equation}
	where $R(\mathbf n_\mu)$ is the three-dimensional rotation matrix, and it is explicitly written as,
	\begin{equation}
		R(\mathbf n_\mu)=\begin{bmatrix}n_\mu^z+\frac{(n_\mu^y)^2}{1+n_\mu^z} & -\frac{n_\mu^x n_\mu^y}{1+n_\mu^z} & n_\mu^x \\ -\frac{n_\mu^x n_\mu^y}{1+n_\mu^z} & n_\mu^z+\frac{(n_\mu^x)^2}{1+n_\mu^z} & n_\mu^y \\ -n_\mu^x & -n_\mu^y & n_\mu^z\end{bmatrix}\equiv \begin{bmatrix} \mathbf u(\mathbf n_\mu) & \mathbf v(\mathbf n_\mu)& \mathbf n_\mu  \end{bmatrix}.
	\end{equation}
	Accordingly the Heisenberg term in the global coordinate $H_{ij\mu\nu}$ transforms in the rotated frames as,
	\begin{equation}
		H_{ij\mu\nu}= \sum_{\alpha,\beta=x,y,z} \tilde{J}_{ij\mu\nu}^{\alpha\beta}\tilde{S}_{i\mu}^\alpha \tilde{S}_{j\nu}^\beta,
	\end{equation}
	where $\tilde{J}_{ij\mu\nu}^{\alpha\beta} \equiv J_{ij\mu\nu} [R(\mathbf n_{\mu})^T R(\mathbf n_{\nu})]^{\alpha\beta} $ represents an ``effective Heisenberg interaction" in the rotated frames, which we explicitly write down the expression as,
	\begin{equation}
		\tilde{J}_{ij\mu\nu}^{\alpha\beta}= J_{ij\mu\nu} \begin{bmatrix} \mathbf u(\mathbf n_\mu)\cdot \mathbf u(\mathbf n_\nu) & \mathbf u(\mathbf n_\mu)\cdot \mathbf v(\mathbf n_\nu) & \mathbf u(\mathbf n_\mu)\cdot \mathbf n_\nu \\ \mathbf v(\mathbf n_\mu)\cdot \mathbf u(\mathbf n_\nu) & \mathbf v(\mathbf n_\mu)\cdot \mathbf v(\mathbf n_\nu) & \mathbf v(\mathbf n_\mu)\cdot \mathbf n_\nu \\
			\mathbf n_\mu \cdot \mathbf u(\mathbf n_\nu) & \mathbf n_\mu \cdot \mathbf v(\mathbf n_\nu) & \mathbf n_\mu \cdot \mathbf n_\nu \end{bmatrix}_{\alpha\beta}.
	\end{equation}
	After the transformation to the local coordinate frame, we perform the standard HP transformation, which is given as,
	\begin{equation}
		\tilde{S}_{i\mu}^x \simeq \frac{\sqrt{2S}}{2} (a_{i\mu}+a_{i\mu}^\dagger), ~ \tilde{S}_{i\mu}^y\simeq \frac{\sqrt{2S}}{2i} (a_{i\mu}-a_{i\mu}^\dagger), ~ \tilde{S}_{i\mu}^z=S-a_{i\mu}^\dagger a_{i\mu}.
	\end{equation} 
	By plugging in the above expression, we expand $H_{ij\mu\nu}$ up to the quadratic order of the magnon Hamiltonian, which is now given as,
	\begin{align}
		H_{ij\mu\nu}& = \tilde{J}_{ij\mu\nu}^{zz}S^2+\sqrt{2}S^{3/2}\bigg[\big(\tilde{J}_{ij\mu\nu}^{xz}-i\tilde{J}_{ij\mu\nu}^{yz}\big)a_{i\mu}+h.c.\bigg] -S\tilde{J}_{ij\mu\nu}^{zz}\big(a_{i\mu}^\dagger a_{i\mu}+a_{j\nu}^\dagger a_{j\nu} \big) \nonumber \\
		& + S\bigg[\frac{1}{2}\big(\tilde{J}_{ij\mu\nu}^{xx}+\tilde{J}_{ij\mu\nu}^{yy}-i\tilde{J}_{ij\mu\nu}^{xy}+i\tilde{J}_{ij\mu\nu}^{yx}\big) a_{i\mu}^\dagger a_{j\nu}+\frac{1}{2}\big(\tilde{J}_{ij\mu\nu}^{xx}-\tilde{J}_{ij\mu\nu}^{yy}+i\tilde{J}_{ij\mu\nu}^{xy}+i\tilde{J}_{ij\mu\nu}^{yx}\big) a_{i\mu}^\dagger a_{j\nu}^\dagger +h.c. \bigg] +\mathcal O (\sqrt{S}). \label{eq:H_expansion}
	\end{align}
	We also rewrite $A_{i\mu}$ in the rotated frames as
	\begin{equation}
		A_{i\mu}= -\sum_{\alpha,\beta=x,y,z} \tilde{D}_{\mu}^{\alpha\beta} \tilde{S}_{i\mu}^\alpha \tilde{S}_{i\mu}^\beta,
	\end{equation}
	where $\tilde{D}_{\mu}^{\alpha\beta} $ represents an ``effective single-ion anisotropy energy" in the rotated frames, which we also write down explicitly as
	\begin{equation}
		\tilde{D}_{\mu\nu}^{\alpha\beta}=D_z\begin{bmatrix} (n_\mu^x)^2 & n_\mu^x n_\mu^y & -n_\mu^x n_\mu^z \\ n_\mu^x n_\mu^y & (n_\mu^y)^2 & -n_\mu^y n_\mu^z \\ -n_\mu^x n_\mu^z & -n_\mu^y n_\mu^z & (n_\mu^z)^2 \end{bmatrix}_{\alpha\beta}.
	\end{equation}
	Using the HP transformation, we also expand $A_{i\mu}$ as
	\begin{align}
		A_{i\mu}& = -\tilde{D}_{\mu}^{zz}S^2 - \sqrt{2}S^{3/2}\bigg[\big(\tilde{D}_{\mu}^{xz}-i\tilde{D}_{\mu}^{yz}\big)a_{i\mu} +h.c.\bigg], \nonumber\\
		& - S\bigg[\big(\tilde{D}_{\mu}^{xx}+\tilde{D}_{\mu}^{yy}-2\tilde{D}_{\mu}^{zz}\big)a_{i\mu}^\dagger a_{i\mu}+\frac{1}{2}\big(\tilde{D}_{\mu}^{xx}-\tilde{D}_{\mu}^{yy}+2i\tilde{D}_{\mu}^{xy}\big)a_{i\mu}^\dagger a_{i\mu}^\dagger +h.c.\bigg]+ \mathcal O(\sqrt{S}). \label{eq:A_expansion}
	\end{align}
	Using Eqs.~(\ref{eq:H_expansion}) and (\ref{eq:A_expansion}), we obtain a magnon Hamiltonian within a linear spin-wave theory as
	\begin{equation}
		H_{\textrm{BdG}} = 2\sum_{i} \sum_{\mu} E_{\mu}a_{i\mu}^\dagger a_{i\mu} + \sum_{i,j} \sum_{\mu,\nu}\bigg[T_{ij\mu\nu}a_{i\mu}^\dagger a_{j\nu} +\Delta_{ij\mu\nu} a_{i\mu}^\dagger a_{j\nu}^\dagger +h.c.\bigg],
	\end{equation}
	where
	\begin{align}
		E_\mu & = -S\sum_{j} \sum_{\nu} \tilde{J}_{ij\mu\nu}^{zz} + S\big(\tilde{D}_{\mu}^{zz}-\tilde{D}_{\mu}^{xx}/2-\tilde{D}_{\mu}^{yy}/2\big), \\
		T_{ij\mu\nu} & = \frac{S}{2}\big(\tilde{J}_{ij\mu\nu}^{xx}+\tilde{J}_{ij\mu\nu}^{yy}-i\tilde{J}_{ij\mu\nu}^{xy}+i\tilde{J}_{ij\mu\nu}^{yx}\big), \\
		\Delta_{ij\mu\nu} & = \frac{S}{2}\big(\tilde{J}_{ij\mu\nu}^{xx}-\tilde{J}_{ij\mu\nu}^{yy}+i\tilde{J}_{ij\mu\nu}^{xy}+i\tilde{J}_{ij\mu\nu}^{yx}\big)-\delta_{ij}\delta_{\mu\nu}\frac{S}{2}\big(\tilde{D}_{\mu}^{xx}-\tilde{D}_{\mu}^{yy}+2i\tilde{D}_{\mu}^{xy}\big).
	\end{align}
	Finally, we arrive the expression of the Bogoliubov Hamiltonian $H_\textrm{BdG}$, which is given as,
	\begin{equation}
		H_{\textrm{BdG}} = \sum_{\mathbf{k}} \sum_{\mu,\nu} \begin{pmatrix} a_{\mathbf k\mu}^\dagger & a_{-\mathbf k\mu}  \end{pmatrix} \begin{pmatrix} \delta_{\mu\nu}E_{\mu} + T_{\mathbf k\mu\nu} & \Delta_{\mathbf k\mu\nu} \\ \Delta^{*}_{\mathbf k\mu\nu} &  \delta_{\mu\nu}E_{\mu} +T_{-\mathbf k\nu\mu} \end{pmatrix} \begin{pmatrix} a_{\mathbf k\nu} \\ a_{-\mathbf k\nu} ^\dagger \end{pmatrix},
	\end{equation}
	where
	\begin{align}
		T_{\mathbf k\mu\nu} & = T_{ii\mu\nu} + \sum_{j}e^{i\mathbf k\cdot(\mathbf r_i-\mathbf r_j)} T_{ij\mu\nu}, \\
		\Delta_{\mathbf k\mu\nu} & = \Delta_{ii\mu\nu} + \sum_{j}e^{i\mathbf k\cdot(\mathbf r_i-\mathbf r_j)} \Delta_{ij\mu\nu}.
	\end{align}
	In general non-collinear order, the BdG Hamiltonian can be diagonalized as, $H_{\textrm{BdG}}=U D U^\dagger$, where $D$ is the diagonal matrix containing the magnon band energy. The bosonic commutation relations of the magnons require the paraunitary condition of the bosonic wave function, which is given as, $U^\dagger\Sigma_z U=\Sigma_z$, where $\Sigma_z$ is the Pauli matrix acting on the particle-hole degree of the freedom. In general, the paraunitary nature of the bosonic wave function makes the deviation of the band topological classifications from the ordinary fermionic topological systems \cite{RevModPhys.88.035005,Ryu_2010}.

	\section{Topological classification of magnon excitations}
	
	\subsection{General consideration}
	
	For a given magnetic ground state, the BdG Hamiltonian of the Holstein-Primakoff magnons are obtained by the method described in the previous section \ref{Appsec:magnon1}. In this section, we consider the general symmetry structure of the magnon Hamiltonian and the magnon band topological classifications. The BdG Hamiltonian of the magnon can be diagonalized as,
	\bea
	H_{\textrm{BdG}}(\mathbf{k})=U_\mathbf{k}D_\mathbf{k}U^\dagger_\mathbf{k},
	\eea 
	where $U_\mathbf{k}$ is the paraunitary matrix that satisfy the paraunitary condition such that $U_\mathbf{k}^\dagger\Sigma_z U_\mathbf{k}=\Sigma_z$ due to the bosonic commutation relation of the magnon, where $\Sigma$ denotes the Pauli matrices acting on the particle-hole space. To obtain the paraunitary wave function, the non-Hermitian Hamiltonian, $\Sigma_{z}H_{\textrm{BdG}} (\mathbf{k})$ is diagonalized\cite{COLPA1978327}. In addition, the particle-hole symmetry of the BdG Hamiltonian, which is given as, \bea
	\Sigma_x H_{\textrm{BdG}}^T(\mathbf{k}) \Sigma_x=H_{\textrm{BdG}}(-\mathbf{k}).
	\eea 
	The particle-hole symmetry of the BdG Hamiltonian allows the decomposition of the wave functions into the positive and negative energy sectors as $U_\mathbf{k}=(V_\mathbf{k}, \Sigma_x V_{-\mathbf{k}})$ where $V_\mathbf{k}$ is $N\times \frac{N}{2}$-dimensional matrix, which is comprised of the eigenvectors with the positive energy \cite{PhysRevB.99.041110,Kondo2020,lu2018magnon}.
	
	\subsection{Collinear order and Dirac magnons}\label{Appsec:Dirac}

	In the case of the collinear magnetic order, the system possesses the additional $U(1)$-spin rotation symmetry($U(1)_c$) along the easy axis. As the pairing term in the BdG Hamiltonian vanishes due to the $U(1)_c$ symmetry, the Bogoliubov-de Gennes (BdG) Hamiltonian can be decoupled into the particle and hole sectors as, 
	\bea
	H_{\textrm{BdG}}(\mathbf{k})=H_{s=+1}\oplus H_{s=-1},
	\eea
	where $H_{s=\pm 1}$ indicates the particle (hole) sector of the magnons carrying the spin $\pm1$ along the collinear axis. When the particle and the hole sectors are decoupled, we can choose the basis of the eigenfunctions, which satisfy the unitary condition, $U_\mathbf{k}^\dagger U_\mathbf{k}=I$. The unitary eigenstates of the magnons follow the same topological classifications of the spinless fermions. In the presence of the space-time inversion symmetry ($\mathcal{P}\mathcal{T}$), where $\mathcal{T}: S_y \rightarrow -S_y$ is the effective time-reversal symmetry of the magnons, ensures $\mathbb{Z}_2$ quantization of the Berry phase along a closed loop in the BZ (AI class in the Altland-Zirnbauer classification). The non-trivial $\pi$-Berry phase along a closed loop topologically protects the odd number of the Dirac cones of the magnonic systems in the interior of the loop. In the commensurate angles of the twisted bilayer systems, the Dirac cones from the top layer and the bottom layers are located at the same point in the folded BZ \cite{PhysRevB.99.195455}. The topological protection of the two individual Dirac cones requires the effective decoupling between the two, which manifests as $U(1)_v$ symmetry.

\end{document}